\documentclass[a4wide]{article}

\usepackage{authblk}

\usepackage[T1]{fontenc}
\usepackage[latin9]{inputenc}
\usepackage{color}
\usepackage{units}
\usepackage{amstext}
\usepackage{amsthm}
\usepackage{amssymb}
\usepackage{amsfonts}
\usepackage{amsmath}
\usepackage[usenames]{xcolor}
\usepackage{graphics}
\usepackage{epstopdf}
\usepackage{graphicx}
\usepackage{tikz,pgf}
\usepackage{pgfplots}
\pgfplotsset{compat=newest}

\usepackage{float}
\usepackage[linesnumbered,ruled]{algorithm2e}

\makeatletter

\floatstyle{ruled}
\newfloat{algorithm}{tbp}{loa}
\providecommand{\algorithmname}{Algorithm}
\floatname{algorithm}{\protect\algorithmname}
\makeatother

\usepackage{esint}

\makeatletter

\newif\ifpreprintversion
\preprintversionfalse

\renewcommand{\S}{\mathcal{S}}
\newcommand{\K}{\mathcal{K}}
\renewcommand{\L}{L}
\newcommand{\df}{\,\mathrm{d}}

\newcommand{\abs}[1]{\left\lvert #1\right\rvert}
\newcommand{\norm}[1]{\left\lVert #1\right\rVert}
\newcommand{\ip}[2]{\left\langle#1,#2\right\rangle}
\newcommand{\R}{\mathbb{R}}
\renewcommand{\div}{\mathop{\mathrm{div}}}

\newcommand{\m}{m}
\newcommand{\N}{\mathbb{N}}
\newcommand{\E}{\mathbb{E}}
\newcommand{\Z}{\mathbb{Z}}

\newcommand{\DS}{\frac{\delta \S}{\delta \rho}}

\newcommand{\ini}{\mathrm{ini}}
\newcommand{\prep}{\mathrm{prep}}
\newcommand{\ana}{\mathrm{analytic}}

\definecolor{jzorange}{RGB}{230, 159, 0}
\definecolor{jzskyblue}{RGB}{86, 180, 233}
\definecolor{jzgreen}{RGB}{0, 158, 115}
\definecolor{jzyellow}{RGB}{240, 228, 66}
\definecolor{jzblue}{RGB}{0, 114, 178}  
\definecolor{jzred}{RGB}{213, 94, 0} 

\makeatother

\begin{document}

\title{Computing diffusivities from particle models out of equilibrium}

\author[$\dagger$]{Peter Embacher}

\author[$\dagger$]{Nicolas Dirr}

\author[$\star$]{Johannes Zimmer}

\author[$\ddag$]{Celia Reina}

\affil[$\dagger$]{School of Mathematics, Cardiff University, Cardiff CF24 4AG, UK}
\affil[$\star$]{Department of Mathematical Sciences, University of Bath, Bath BA2 7AY, UK}
\affil[$\ddag$]{Department of Mechanical Engineering and Applied Mechanics, University of Pennsylvania, Philadelphia PA 19104, USA }

\maketitle

\begin{abstract}
  A new method is proposed to numerically extract the diffusivity of a (typically nonlinear) diffusion equation from underlying
  stochastic particle systems.  The proposed strategy requires the system to be in local equilibrium and have Gaussian
  fluctuations but is otherwise allowed to undergo arbitrary out of equilibrium evolutions. This could be potentially relevant for
  particle data obtained from experimental applications.  The key idea underlying the method is that finite, yet large, particle
  systems formally obey stochastic partial differential equations of gradient flow type satisfying a fluctuation-dissipation
  relation.  The strategy is here applied to three classic particle models, namely independent random walkers, a zero range
  process and a symmetric simple exclusion process in one space dimension, to allow the comparison with analytic solutions.
\end{abstract}

\section{Introduction}
\label{sec:Introduction}

Diffusive processes are ubiquitous in natural and man-made materials and devices, ranging from mass transport in cells over ion
diffusion in batteries to diffusion of pollutants in oceans and the atmosphere. They can often be described on a fine scale via
particles, and on a coarser (continuum) scale by partial differential equations. While the former typically provides higher
physical fidelity, the computational efficiency of the latter enables us to reach the length and time scales required in many
applications.

Yet, simulations at the continuum level require the knowledge of correct material parameters or transport coefficients, such as
the diffusivity for mass transport processes.  These can be determined from lower scale models or experimental observations, which
is precisely the problem we study here. In particular, we consider diffusive systems described by evolution equations of the form
\begin{equation*}
\partial_t\rho = \div \left( D(\rho) \nabla \rho \right),
\end{equation*}
where $\rho = \rho(t,x)$ is the density and $D(\rho)$ is the diffusion coefficient.
 
Existing methods to compute diffusivities include equilibrium strategies based on linear response theory, e.g., Green-Kubo or mean
square displacement methods (see, for example,~\cite[Section 4.4.1]{Frenkel2002a} and~\cite{Keffer2005a}), and non-equilibrium
molecular dynamic techniques (see, for example,~\cite[Chapters 2 and 8]{Allen1989a},~\cite[Chapter 13]{Tuckerman2010a}
or~\cite{Evans2007a}). Although these methods have been proven very successful, they require the system to be in special
configurations, such as macroscopic equilibrium or steady state, or the simulation of modified equations of motion.  This may pose
challenges for a range of problems where transport coefficients should be inferred from experimental data, where the special
configurations required by these methods might not always be easily achievable experimentally.

In this article, we develop a new computational strategy to determine diffusivity coefficients from quite general non-equilibrium
particle evolutions: the primary assumptions are local equilibrium and Gaussian nature of the fluctuations of the density (see
Subsection~\ref{sec:Key-assumpt-meth} for further details). The key idea is that the evolution of a finite, yet large, number of
particles can formally be described by a stochastic partial differential equation (SPDE) satisfying a fluctuation-dissipation
relation. In particular, when the deterministic part of the SPDE is written as a gradient flow of the entropy functional, the
density evolves in the direction of steepest ascent of the entropy (while compatible with the constraint of conservation of mass)
at a speed that is characterized by a mobility coefficient~\cite{Jordan1998a,Reina2015a}.  This mobility is also encoded in the
noise term of the SPDE, by the fluctuation-dissipation relation, and it may be directly related to the diffusivity through the
knowledge of the entropy.  We extend in this study a classic approach for the computation of the noise term of a stochastic
ordinary differential equation to the infinite-dimensional setting of SPDEs, which allows us to compute the mobility from the
fluctuations observed in particle evolution data.

There are various interesting features of this approach. Firstly, the particle evolution data is allowed to be in or out of
equilibrium, which provides great flexibility for the input data that can be considered. Furthermore, it allows us to recover the
function $D(\rho)$ within the range of densities covered by the particle data --- we chose here a sinusoidal shape for the initial
density to demonstrate this feature. This is in contrast to common equilibrium techniques, which only deliver the diffusivity at
the simulated density.  Secondly, the particles' mean square displacement (MSD) is not assumed to be linear in time, as it would
be in conventional calculations of the diffusivity using Einstein's relation. Rather, the diffusivity is directly computed from
the fluctuations, without assuming a specific form of the temporal evolution. This versatility of the method is demonstrated by
considering a symmetric nearest neighbour exclusion process in one space dimension, for which MSD $\sim t^{1/2}$, thus rendering
the conventional method inapplicable; the method determines the diffusivity correctly for this case. Lastly, we note that,
although we restrict the analysis to nonlinear diffusion problems, which in itself exhibit a wide variety of interesting phenomena
and applications, the potential applicability is much larger. Indeed, we only use a gradient flow structure of the thermodynamic
evolution and the fluctuation-dissipation relation; this setting comprises a much wider range of dissipative
phenomena~\cite{Ottinger2005a,Mielke2011b}.

In this article we demonstrate the methodology for stochastic particle problems on lattices with analytic solutions to their
macroscopic evolution equation, so that the errors can be exactly determined.  In particular, we study independent random walkers,
a zero range process with quadratic jump rates and a symmetric simple exclusion process. We could not find the analytic
description of the continuum limit for the second of these processes in the literature, and the derivation of the analytic
expression is sketched in Subsection~\ref{sec:Zero-range-process}. In all of these cases, the input to the method is obtained from
kinetic Monte Carlo simulations of these processes. However, the input data can also be, in principle, obtained from experiments;
the applicability of the method in experimental settings will be investigated in future studies.

The structure of the paper is as follows. In Section~\ref{sec:Therdm-metr-theory}, we describe the thermodynamic formulation of
diffusion processes, both in deterministic and stochastic form, in Subsection~\ref{sec:Diff-proc-therm}, and develop the
computational strategy in Subsection~\ref{sec:Comp-strat}. Section~\ref{sec:Numer-impl} provides details on the numerical
implementation. Next, in Section~\ref{sec:Part-models-stud} we describe the particle processes used as test cases, and present the
computational results in Section~\ref{sec:Comp-results}. Finally, some conclusions are drawn in
Section~\ref{sec:Conclusions-outlook}, where also an outlook to an application of the proposed method to particles experiencing
Kawasaki dynamics is given.

\section{Thermodynamics metric: theory and computational method}
\label{sec:Therdm-metr-theory}

\subsection{Diffusive processes and fluctuating hydrodynamics: thermodynamic entropy and metric}
\label{sec:Diff-proc-therm}

As sketched in the introduction, we consider diffusive systems of the form
\begin{equation}
  \label{eq:diff-linear-D}
  \partial_t \rho = \div (D(\rho) \nabla \rho)
\end{equation}
(complemented with initial and boundary conditions) and develop a computational strategy to determine the \emph{diffusivity}
$D(\rho)$ from non-equilibrium evolutions of the underlying particle models.

The proposed methodology starts by reformulating the differential equation~\eqref{eq:diff-linear-D} in a form that reveals the
thermodynamic nature of the equation. Namely, equation~\eqref{eq:diff-linear-D} can be cast as
\begin{equation}
  \label{eq:diff-linear-GF}
  \partial_t \rho = \div (D(\rho) \nabla \rho) = - \div\left(m(\rho) \nabla \DS (\rho)\right) , 
\end{equation}
where $\DS$ is the variational derivative of the entropy $\S(\rho) = \int s(\rho(x)) \df x$, $\left(\DS \right)'=s''(\rho)$
represents its derivative with respect to $\rho$ and
\begin{equation}
  \label{eq:d-and-m}
  \m(\rho) = - \frac {D(\rho)}{\left( \DS \right)'} \geq 0
\end{equation}
is the mobility. As an example, the linear diffusion problem (i.e., constant $D(\rho)\equiv D$), satisfies
equation~\eqref{eq:diff-linear-GF} with the \emph{Boltzmann entropy} $\S(\rho) = - \int \rho \log(\rho) \df x$ (in dimensionless
form) and mobility $m(\rho)=D \rho$.  Although the second equality in~\eqref{eq:diff-linear-GF} is straightforward to verify
from~\eqref{eq:d-and-m}, the associated thermodynamic formulation of the evolution,
$\partial_t \rho = - \div\left(m(\rho) \nabla \DS (\rho)\right) =: \K \DS$, has a much deeper meaning. Specifically, the operator
$\K(\rho) \xi = -\div(D \rho \nabla \xi)$ defines a geometry, the so-called \emph{Wasserstein geometry}, in which the entropy $\S$
experiences a steepest ascent.  This fundamental insight by Jordan, Kinderlehrer and Otto~\cite{Jordan1998a} has triggered much
activity in the past two decades; we sketch a few key results in Appendix~\ref{sec:Entr-grad-flow}.

Equation~\eqref{eq:diff-linear-GF} arises as the limit, the so-called hydrodynamic limit, of infinitely many particles under
parabolic scaling of space and time. We refer the reader to Sections~\ref{sec:Numer-impl} and~\ref{sec:Part-models-stud} for a
concise description of this limit in the context of lattice systems. Key to the proposed method is that the evolution of the
density $\rho^L$ describing the evolution of large, yet \emph{finite} number of particles is formally given by a stochastic
partial differential equation. For instance, the motion of finitely many random walkers satisfies approximately the equation
\begin{align}
  \label{eq:dean}
  \partial_t \rho^\L &= \div (D \nabla \rho^\L) + \frac 1 {\sqrt{\L^d}} \div(\sqrt{2 D\rho^\L} \dot{W}_{x,t}) , 
\end{align}
where the diffusivity $D$ depends on the jump rate of the walkers, while in the limit of infinite number of particles the density
satisfies~\eqref{eq:diff-linear-GF}. Here, $\dot{W}_{x,t}$ is a space-time white noise, $1/L$ is the lattice spacing and $d$ is
the dimension of space. We note that~\eqref{eq:dean}, which is also satisfied for the collective motion of finitely many Brownian
particles~\cite{Dean1996a}, is an example of \emph{fluctuating hydrodynamics}~\cite{Eyink1996a, Jack2014a,Hurtado2013a}, see
also~\cite{Renesse2009a} for a different derivation. The existence of a solution is an open mathematical problem, even for
constant $D$, though equations of fluctuating hydrodynamics are widely used for simulations.

In general, the fluctuations of finitely many particles around the hydrodynamic limit, cf.~\eqref{eq:diff-linear-GF}, are
therefore described by a stochastic partial differential equation (SPDE) and they encode the diffusivity $D$ that we wish to
identify. More precisely, as will be described in the next subsection, the fluctuations are directly related to the mobility
$m(\rho)$, from which $D(\rho)$ can be computed by means of equation~\eqref{eq:d-and-m}. For the processes we consider, the
entropy $\S$ is well-known. Yet, more generally, only entropy differences $\delta \S$ are additionally required and these can be
computed with standard techniques~\cite[Chapter 7]{Frenkel2002a}.

We remark that the proposed method is in principle much more widely applicable, i.e., beyond the realm of diffusion, and this
extension to a wider class of problems will be sketched in the conclusions in Section~\ref{sec:Conclusions-outlook}.

\subsubsection{Key assumptions of the method} 
\label{sec:Key-assumpt-meth}

The method requires three key ingredients: a diffusive stochastic particle model, Gaussian fluctuations of the density and local
equilibrium.  Roughly speaking, local equilibrium can be understood as follows.  In equilibrium, the probability distribution of
the particles defines a unique invariant measure for each macroscopic density and discretisation level $L$. For fixed $L$, one can
thus think of a family of invariant measures parametrised by the total mass, or equivalently by the density $\rho$. Then, assuming
that we know the associated macroscopic density $\rho (t,x)$ at each $x$ at a time $t$, we define for the given discretisation
level $L$ a measure by piecing together the invariant measures defined in equilibrium corresponding to the value of $\rho$. The
resulting measure is itself not invariant; yet one expects it to be ``almost invariant'', in the sense that its evolution under
the adjoint of the generator of the process does not vanish, but is small in a well-controlled way. For the precise definition of
local equilibrium, we refer to the book of Kipnis and Landim~\cite[Chapter 3]{Kipnis1999a}.

\subsection{Computational strategy}
\label{sec:Comp-strat}

We now consider particle processes leading, in general, to a nonlinear diffusion as their hydrodynamic limit. We rewrite these
equations as
\begin{equation}
  \label{eq:SPDE-det}
  \partial_t\rho  = \div (D(\rho) \nabla\rho) = -\div \left(\m(\rho) \nabla \DS\right) = \frac{1}{2}\Delta(\Phi(\rho)),
\end{equation}
where $\Phi'(\rho)/2 = D(\rho)$ as defined by the equality above. The density for a large but finite number of particles,
$\rho^\L$, can be approximated to leading order by the hydrodynamic limit given in~\eqref{eq:SPDE-det}.  The next order, the
fluctuations, can be measured via $Y^\L(t,x):= \sqrt{\L^d} \left(\rho^\L(t,x)-\rho(t,x) \right)$, where the scaling $\sqrt{\L^d}$
guarantees a finite non-zero value as $\L\to \infty$. In this limit, the fluctuations solve a linear stochastic partial
differential equation (see~\cite{Ferrari1988a} for the zero range process discussed in Subsection~\ref{sec:Zero-range-process}
and~\cite{Landim2008a} for the simple exclusion process of Subsection~\ref{sec:Simple-excl-proc}). This therefore allows to
formally approximate the evolution of $\rho^\L$ by the SPDE
\begin{equation}
  \label{eq:dean-strategy}
   \partial_t\rho^\L  = \frac{1}{2}\Delta(\Phi(\rho^\L)) +\frac{1}{\sqrt{\L^d}}
   \div \left(\sqrt{2\m(\rho^\L)}\dot{W}_{x,t} \right). 
\end{equation}
We further define the weak form of the fluctuations as
\begin{equation}
  \label{eq:def_Y-N}
 Y^\L_\gamma(t) := \sqrt{\L^d} \ip{\gamma}{\rho^\L(t,\cdot)-\rho(t,\cdot)},
\end{equation}
where $\gamma \in C^2_0(\Omega, \R)$ is a test function and $\ip{}{}$ denotes the $L^2$ inner product defined on the domain
$\Omega$. For our purposes, $\gamma$ will be chosen to have local support so as to measure the fluctuations in the neighbourhood
of a given point $x_0\in \Omega$.

The limit of the stochastic processes $Y^\L$ and $Y^\L_{\gamma}$ are here denoted $Y$ and $Y_{\gamma}$, respectively, and they
satisfy $Y_{\gamma}=\ip{Y}{\gamma}$. Formally, $Y^\L_{\gamma}=\langle Y,\gamma\rangle+O(1/\L^d)=Y_\gamma +O(1/\L^d)$, where $Y$
and $Y_{\gamma}$ obey the Ornstein-Uhlenbeck processes defined by
\begin{align}
  \df Y &=\frac{1}{2}\Delta( \Phi'(\rho)Y) \df t +\div(\sqrt{2\m(\rho)} \df W_{x,t})  \label{eq:fluctuation} \\
  \intertext{and} 
  \df Y_\gamma(t) & =\frac{1}{2}\left\langle \Delta\gamma,\Phi'(\rho(t,.)) Y(t,\cdot) \right
  \rangle \df t -\left\langle \nabla\gamma,\sqrt{2\m(\rho(t,\cdot))} \df W_{x,t} \right\rangle, \label{eq:def_Y} 
\end{align}
respectively.

Structurally,~\eqref{eq:def_Y} is an infinite-dimensional analogue of the finite-dimensional stochastic ordinary differential
equation
\begin{equation}
  \label{eq:sde}
  \mathrm{d} X= f \df t  +\sqrt{\sigma} \df W ,
\end{equation}
for which one can easily compute the strength of the noise $\sigma$ as
\begin{equation}
  \label{eq:formula-sigma}
  \lim_{h\searrow 0}\frac{1}{h}\E\left[\left[X(t_0+h)-X(t_0)\right]^2\right]= d \sigma,
\end{equation}
where $\lim_{h\searrow 0}$ indicates the limit of $h$ to $0$ from above. Here the left-hand side can be approximated by computer
simulations.

We claim that an analogous statement holds for the infinite-dimensional case~\eqref{eq:def_Y} as well, namely
\begin{equation}
  \label{eq:formula-Y-sigma}
   \frac{1}{2 h} \mathbb{E}\left[\left[Y_\gamma(t_0+h)-Y_\gamma(t_0)\right]^{2}\right] =
  \ip{\m(\rho(t_0,\cdot)) \nabla\gamma}{\nabla\gamma},
\end{equation}
where a sufficiently localised function $\gamma$ around a given point $x_0 \in \Omega$, delivers an approximation of
$\m(\rho(t_0,x_0))$ as
\begin{equation}
  \label{Eq:m_approx}
  m(\rho(t_0,x_0)) \simeq \frac{ \lim_{h\searrow 0}\frac{1}{h}
    \mathbb{E}\left[\left[Y_\gamma(t_0+h)-Y_\gamma(t_0)\right]^{2}\right]}{2\ip{\nabla\gamma}{\nabla\gamma}}.
\end{equation}
This relation will allow us to extract $\m$ and hence the diffusivity $D$, via~\eqref{eq:d-and-m}, from the fluctuations of the
system, for large enough particle numbers.  Further details on its numerical implementation will be given in
Section~\ref{sec:Numer-impl}.

To establish~\eqref{eq:formula-Y-sigma}, we separate the so-called quadratic variation of the process~\eqref{eq:def_Y} from the
rest (this is a standard problem in financial mathematics, see, e.g., \cite{Barndorff-Nielsen2002a}). More precisely, we consider
a new random variable $F(Y_\gamma)$, with $F\in C^{2}(\R,\R)$.  By It\^{o}'s formula (see~\cite[Chapter 4]{Oksendal2003a}), this
new variable satisfies
\begin{align}
  \label{eq:dF(Y)}
  \df F (Y_\gamma)(t) & =  F'(Y_\gamma(t)) \df Y_\gamma(t)+\frac{F''(Y_\gamma(t))}{2}
  \ip{ 2\m(\rho(t_0,\cdot)) \nabla\gamma}{\nabla\gamma} \df t , 
\end{align}
where for the last term of~\eqref{eq:def_Y} we have made use of the fact that $\dot{W}_{x,t}$ is a space-time white noise.  We
choose
\begin{equation}
  \label{eq:choice-F}
  F(x):=(x-Y_\gamma(t_0))^{2} ,
\end{equation}
where $t_0\geq0$ is an arbitrary initial time and write (having in mind the left-hand side of~\eqref{eq:formula-Y-sigma})
\begin{equation}
  \mathbb{E}\left[\left[Y_\gamma(t_0+h)-Y_\gamma(t_0)\right]^{2}\right]
  = 
  \mathbb{E}\left[F(Y_{\gamma}(t_0+h)) \right]
  =
  \mathbb{E}\left[ \int_{t_0}^{t_0+h} \df F(Y_{\gamma}(t)) \right]. 
\end{equation}
Using equations~\eqref{eq:dF(Y)}--\eqref{eq:choice-F}, we can rewrite this identity as
\begin{equation}
 \label{eq:wholetermb}
  \begin{split}
    & \mathbb{E}\left[\left[Y_\gamma( t_0+h)-Y_\gamma(t_0)\right]^{2}\right] \\
    &=
    \E\left[\int_{ t_0}^{ t_0+h}
      F'(Y_{ \gamma}(t)) \df Y_{ \gamma}(t)\right]
    + 
    \frac{1}{2}\E\left[\int_{t_0}^{t_0+h} F''(Y_{ \gamma}(t))  \ip{ 2\m(\rho(t_0,\cdot)) \nabla\gamma}{\nabla\gamma} \df t \right] \\
    &=  
    2\E\left[\int_{t_0}^{t_0+h}
      (Y_{ \gamma}(t)-Y_{\gamma}(t_0)) \df Y_{ \gamma}(t)\right] +   \int_{t_0}^{t_0+h} \ip{ 2 \m(\rho(t_0,\cdot)) 
      \nabla\gamma}{\nabla\gamma} \df t. 
  \end{split}
\end{equation}
Hence
\begin{multline}
  \label{eq:z-dot}
  \lim_{h\searrow 0}
  \frac{1}{h} \mathbb{E}\left[\left[Y_\gamma( t_0+h)-Y_\gamma(t_0)\right]^{2}\right] \\
  =  
  \lim_{h\searrow 0} 
  \frac{2}{h}\E\left[\int_{t_0}^{t_0+h}
    (Y_{ \gamma}(t)-Y_{\gamma}(t_0)) \df Y_{ \gamma}(t)\right] +    \ip{ 2\m(\rho(t_0,\cdot)) \nabla\gamma}{\nabla\gamma}.
\end{multline}
To prove~\eqref{eq:formula-Y-sigma}, we show that the first term on the right-hand side vanishes,
\begin{equation}
  \label{eq:firstterm}
  \lim_{h\searrow 0}\frac{2}{h} \mathbb{E}\left[ \int_{t_0}^{t_0+h}  (Y_{ \gamma}(t)-Y_{ \gamma}(t_0)) \df Y_{ \gamma}(t) \right] =0 . 
\end{equation} 
This results follows from~\eqref{eq:def_Y} and H\"{o}lder's inequality followed by Young's inequality, namely 
\begin{equation}
  \label{eq:product}
  \begin{split}
    \ & \E\left[\int_{t_0}^{t_0+h} (Y_{\gamma}(t)-Y_{ \gamma}(t_0)) \df Y_{\gamma}(t) \right] \\
    & =\E\left[\int_{t_0}^{t_0+h} \left( Y_{\gamma}(t) -Y_{\gamma}(t_0) \right)
      \ip{\Delta\gamma}{\frac{1}{2}\Phi' (\rho(t,\cdot)) Y(t,\cdot)} \df t\right]\\  
   & \leq \sqrt{\E\left[\int_{t_0}^{t_0+h}\left( Y_{ \gamma}(t) -Y_{\gamma}(t_0) \right)^{2}\df t \right] 
      \cdot\E \left[ \int_{t_0}^{t_0+h}\ip{\Delta\gamma}{\frac{1}{2}\Phi' (\rho(t,\cdot)) Y(t,\cdot)}^{2}\df t\right]}\\
  & = \sqrt{\int_{t_0}^{t_0+h}\E\left[\left( Y_{ \gamma}(t) -Y_{\gamma}(t_0) \right)^{2}\right]\df t  
      \cdot \int_{t_0}^{t_0+h}\E \left[\ip{\Delta\gamma}{\frac{1}{2}\Phi' (\rho(t,\cdot)) Y(t,\cdot)}^{2}\right]\df t} \\
& \le
\frac{1}{2}\int_{t_0}^{t_0+h}\E\left[\left( Y_{ \gamma}(t) -Y_{\gamma}(t_0) \right)^{2} \right] \df t+
\frac{1}{2}\int_{t_0}^{t_0+h}\E\left[\ip{\Delta\gamma}{\frac{1}{2}\Phi' (\rho(t,\cdot)) Y(t,\cdot)}^{2}\right]\df t.
\end{split}
\end{equation}
We can see from~\eqref{eq:wholetermb} and~\eqref{eq:product} that with
\begin{align*}
  Z(t) &:= \int_{t_0}^t \E\left[\left[ Y_{ \gamma}(s) -Y_{\gamma}(t_0) \right]^{2} \right]\df s \\
  \intertext{and} 
  R\left(t\right) &:= \int_{t_0}^{t}\E\left[\ip{\Delta\gamma}
    {\frac{1}{2}\Phi' (\rho(t,\cdot)) Y(t,\cdot)}^{2}\right]\df t
  + \int_{t_0}^{t}  \ip{ 2\m(\rho(t_0,\cdot)) \nabla\gamma}{\nabla\gamma} \df t 
\end{align*}
it holds that
\begin{equation*}
  \dot Z(t)\le Z(t)+R(t),
\end{equation*}
where $R$ is bounded and continuous. Thus, by Gronwall's lemma (e.g.,~\cite[Lemma 4.1.2]{Guckenheimer1990a})
\begin{equation*}
  Z(t)\le e^{(t-t_0)}\int_{t_0}^te^{-(s-t_0)}R(s) \df s, 
\end{equation*}
and hence $Z(t_0+h)= O(h^2)$, as $R(t_{0}+h)=O(h)$. Inserting this in the second but last line in~\eqref{eq:product}, we find that
\begin{equation*}
  \mathbb{E}\left[ \int_{t_0}^{t_0+h} (Y_{ \gamma}(t)-Y_{ \gamma}(t_0)) \df Y_{ \gamma}(t) \right]= O(h^{\frac{3}{2}}),
\end{equation*}
since the first product under the square root is $O(h^2)$ and the second is $O(h)$.
Thus~\eqref{eq:firstterm} is established.

\section{Numerical implementation}
\label{sec:Numer-impl}

We first describe the general particle setting studied here (specific examples will be described in detail in
Section~\ref{sec:Part-models-stud}), and then discuss the numerical implementation for this class of systems.

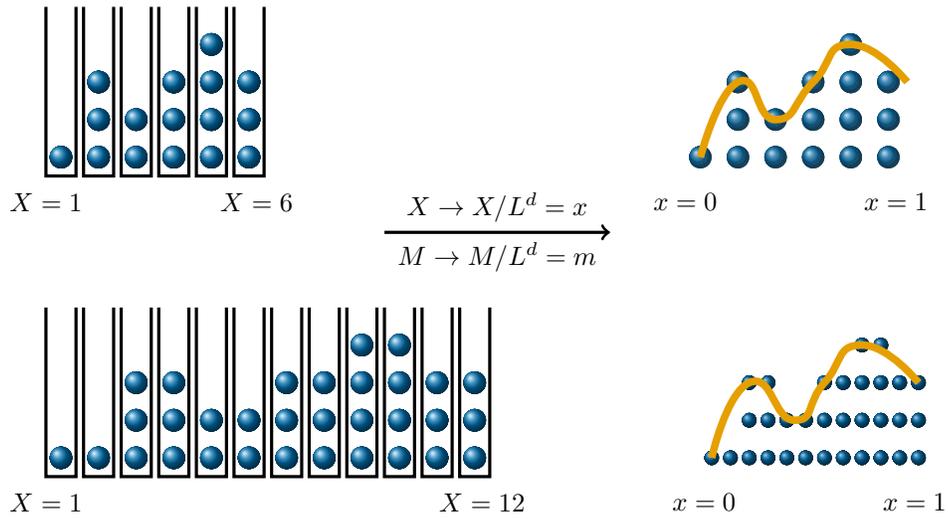
\begin{figure}
\begin{tikzpicture}[scale=0.5]

\def\leftshift{10}
 \foreach \x in {1, ...,6}
 \draw [very thick] (\x-\leftshift, 4) -- (\x-\leftshift,-.5) -- (\x+.8-\leftshift,-.5) -- (\x+.8-\leftshift,4) ;
 \def\numBitsList{0, 0, 2,1,2,3,2}
 \def\numBitsArray{{\numBitsList}}
 \foreach \x  in {1,...,6} {
   \pgfmathparse{\numBitsArray[\x]} \let\lastball\pgfmathresult
   \foreach \y in {0,...,\lastball} {
     \shade[ball color=jzblue]  (\x-\leftshift+.4,\y) circle (0.3cm) ; 
 } }
 
 \node at (-\leftshift+1,-.7) [below] {$X=1$};
 \node at (-\leftshift+6.6,-.7) [below] {$X=6$};
 
\def\leftshift{-7}
 \foreach \x in {1, ...,6} {
 \def\numBitsList{0, 0, 2,1,2,3,2}
 \def\numBitsArray{{\numBitsList}}
 \foreach \x  in {1,...,6} {
   \pgfmathparse{\numBitsArray[\x]} \let\lastball\pgfmathresult
   \foreach \y in {0,...,\lastball} {
     \shade[ball color=jzblue, opacity = 0.
05]  (\x-\leftshift+.4,\y) circle (0.3cm) ; 
 } } }
\draw [jzorange, line width = 1mm] plot [smooth, tension=1] coordinates { (1-\leftshift+.4,0) (2-\leftshift+.4,2) (3-\leftshift+.4,1) 
  (4-\leftshift+.4,2) (5-\leftshift+.4,3) (6.5-\leftshift+.4,2)};

 \node at (-\leftshift+1,-.7) [below] {$x=0$};
 \node at (-\leftshift+6.6,-.7) [below] {$x=1$};

\def\leftshift{-10}
\def\downshift{8}
 \foreach \x in {1,...,12}
 \draw [very thick] (\x+\leftshift, -4) -- (\x+\leftshift,-8.5) -- (\x+.8+\leftshift,-8.5) -- (\x+.8+\leftshift,-4) ;
 \def\numBitsList{0, 0, 2,1,2,3,2}
 \def\numBitsArray{{\numBitsList}}
 \foreach \x  in {1,...,6} {
   \pgfmathparse{\numBitsArray[\x]} \let\lastball\pgfmathresult
   \foreach \y in {0,...,\lastball} {
     \shade[ball color=jzblue]  (2*\x+\leftshift+.4,\y-8) circle (0.3cm) ;
        \shade[ball color=jzblue]  (2*\x-1+\leftshift+.4,\y-8) circle (0.3cm) ; 
 } }

 \node at (\leftshift+1,-.7-\downshift) [below] {$X=1$};
 \node at (\leftshift+12.6,-.7-\downshift) [below] {$X=12$};
 
 \def\leftshift{8}
 \foreach \x in {1, ...,12} {
 \def\numBitsList{0, 0, 2,1,2,3,2}
 \def\numBitsArray{{\numBitsList}}
 \foreach \x  in {1,...,6} {
   \pgfmathparse{\numBitsArray[\x]} \let\lastball\pgfmathresult
   \foreach \y in {0,...,\lastball} {
         \shade[ball color=jzblue, opacity = 0.02]  (\x+\leftshift+.2,\y-8) circle (0.2cm) ;
        \shade[ball color=jzblue, opacity = 0.02]  (\x-.5+\leftshift+.2,\y-8) circle (0.2cm) ; 
 } } }
 \def\leftshift{-7.5}
\draw [jzorange, line width = 1mm] plot [smooth, tension=1] coordinates { (1-\leftshift+.2,-\downshift) (2-\leftshift+.2,2-\downshift) (3-\leftshift+.4,1-\downshift) 
  (4-\leftshift+.2,2-\downshift) (5-\leftshift+.2,3-\downshift) (6.5-\leftshift+.2,2-\downshift)};

 \node at (-\leftshift+1,-.7-\downshift) [below] {$x=0$};
 \node at (-\leftshift+6.6,-.7-\downshift) [below] {$x=1$};

\draw [very thick, ->] (0,-2) -- node [above,pos=0.5] {$X \to X/L^d = x$} node [below,pos=0.5] {$M \to M/L^d = m$}  (6,-2) ; 
\end{tikzpicture}
\caption{Visualisation of the hydrodynamic limit procedure in microscopic coordinates $X \in \Lambda$ (left panel) and the
  macroscopic space $x\in \Omega$ (right panel); here the space dimension is $d=1$. The top figures correspond to $L=6$ and the
  bottom ones to ${L=12}$.  The particles on the left have mass $M =1$, and in the hydrodynamic limit procedure space and mass are
  both rescaled by a factor $1/L^d$, giving rise to the macroscopic space variable $x$ and macroscopic mass $m$. Thus, all four
  configurations have the same density and the total macroscopic mass is kept constant. We note that the balls on the right are
  imaginary and plotted only to guide the construction of the limit function $\rho(t,x)$, represented with a solid line.}
  \label{fig:hydro}
\end{figure}

We always consider $N$ particles distributed on a periodic lattice $\Lambda = \Z^d/(\L\Z)^d$, that is, the torus in $\Z^d$ of
length $\L$ in each direction. We denote lattice coordinates by capital Latin letters, while $\eta$ stands for a lattice
configuration. Thus, $\eta(T, X)$ is the number of particles at site $X \in \Lambda$ and time $T$. Together with the particle
systems, we consider their hydrodynamic limit (both the stochastic/fluctuating form~\eqref{eq:dean-strategy} and the deterministic
limit~\eqref{eq:SPDE-det}), which is on what we call the \emph{macroscopic} or \emph{continuum} scale. The macroscopic spatial and
temporal coordinates will be denoted by $x$ and $t$, respectively, in accordance with the notation used for the partial
differential equations in Section~\ref{sec:Therdm-metr-theory}.  These coordinates are related to the \emph{microscopic}
coordinates $X$ and $T$ via the parabolic scaling, $x = X/\L$ and $t = T/\L^2$, thus $x \in \Omega := (0,1)^d$, the
$d$-dimensional unit cube. The limit passage $N\to\infty$ and $\L \to \infty$ is such that $N/\L^d$ is kept constant.  Also, the
microscopic mass of each particle is rescaled in the limit procedure by $1/L^d$, thus keeping the total mass at the macroscopic
scale constant, see Figure~\ref{fig:hydro}. This interpretation endows the empirical measure $\rho^\L$ with the physical meaning
of the density of the system.

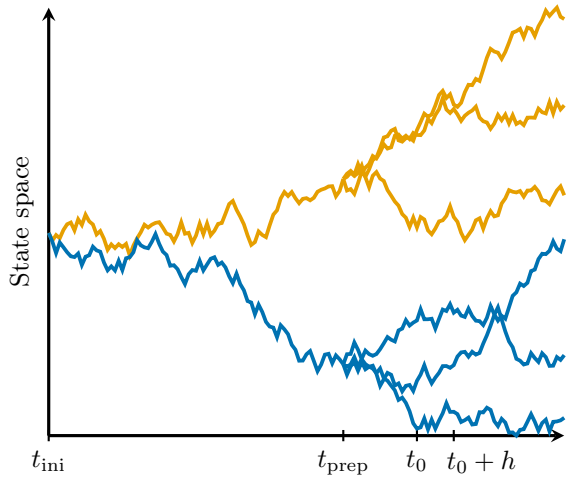
\begin{figure}
\center{
  \begin{tikzpicture}[scale=1]
    \begin{axis}[
      ylabel=State space,
      grid=none,
      xtick=\empty,
      ytick=\empty,
      axis x line = left,
      axis y line = left, 
      axis line style = very thick, 
      extra x ticks={0, 80, 100, 110},
      extra x tick labels={$t_\ini$, $t_\prep$, $t_0$, \phantom{Ham}$t_0+h$},
      clip=false,
      no marks,
      legend style = {draw = none},
      every tick/.style={
        black,
        thick, 
      },
]
      \addplot [line width=.5mm, solid,  color=jzorange] table [x=timestep, y=Traj1, col sep=comma] {visualisationoftimes2early.csv};
      \addplot [line width=.5mm, solid,  color=jzorange] table [x=timestep, y=Traj11, col sep=comma] {visualisationoftimes2late.csv};
      \addplot [line width=.5mm, solid,  color=jzorange] table [x=timestep, y=Traj12, col sep=comma] {visualisationoftimes2late.csv};
      \addplot [line width=.5mm, solid,  color=jzorange] table [x=timestep, y=Traj13, col sep=comma] {visualisationoftimes2late.csv};
      
      \addplot [line width=.5mm, solid,  color=jzblue] table [x=timestep, y=Traj2, col sep=comma] {visualisationoftimes2early.csv};
      \addplot [line width=.5mm, solid,  color=jzblue] table [x=timestep, y=Traj21, col sep=comma] {visualisationoftimes2late.csv};
      \addplot [line width=.5mm, solid,  color=jzblue] table [x=timestep, y=Traj22, col sep=comma] {visualisationoftimes2late.csv};
      \addplot [line width=.5mm, solid,  color=jzblue] table [x=timestep, y=Traj23, col sep=comma] {visualisationoftimes2late.csv};
  
    \end{axis}
\end{tikzpicture}
\caption{Structure of the $R$ realisations (for simplicity symbolised as scalar) as a function of time. Between $t=t_{\ini}$ and
  $t = t_{\prep}$, $R_1$ samples are simulated, and each of them gives rise to $R_2$ realisations from time $t_{\prep}$ on.  Thus
  in total there are $R = R_1 \cdot R_2$ realisations, which are evaluated in the time interval $[t_0,t_0+h]$. In the sketch of
  the figure $R_1 = 2$ and $R_2=3$.}
  \label{fig:R}}
\end{figure}

With this notation regarding the micro- and macro-scale, we are now ready to provide further details on the implementation
of~\eqref{Eq:m_approx}. In particular, the integrals and derivatives in such equation are approximated by finite differences on
the lattice scale. The deterministic states $\rho$ are substituted by averages over $R$ realisations of their stochastic
counterparts and approximated in the following fashion. At the rescaled positions $x = \frac{X}{\L}$ with
$X \in \{1, \dots, L\}^d$, the density $\rho$ is approximated as
\begin{equation}
  \label{eq:rho-eta}
  \rho(t,x) \approx \frac{1}{R} \sum_{r=1}^{R} \eta_r(t \L^2, x \L) . 
\end{equation}
All these realisations originate from an initial configuration $\eta(t_{\text{ini}}L^2)$ at time $t_{\text{ini}} < t_0$, which is
set up beforehand. Starting from such a configuration, one typically first needs to overcome a transient regime, before a local
equilibrium is reached. In computations, we therefore wait for a relatively long time $t_{0} -t_{\ini}$ and then start the actual
measurement. As this waiting time is relatively costly from a computational perspective and many realisations $R$ are needed for
accurate estimates of the expectations, the following compromise is made, visualised in Figure~\ref{fig:R}. We choose a time
$t_{\prep}$ with $t_{\prep} < t_0$ and $t_0 - t_{\prep} \ll t_0-t_{\ini}$ sufficient for attaining local equilibrium. Then,
between ${t=t_{\ini}}$ and $t = t_{\prep}$, $R_1$ samples are simulated. For each of these $R_1$ realisations, $R_2$ realisations
are launched at time $t = t_{\prep}$ from the data obtained at $t_{\prep}$. This procedure gives rise to a total of $R = R_1 R_2$
random initial conditions at $t=t_0$, which are all associated to the same macroscopic state.  In the subsequent evaluations of
the trajectories in the time period $[t_0,t_0+h]$ all $R$, realisations are treated equally.
  
\begin{figure}
  \centerline{\input{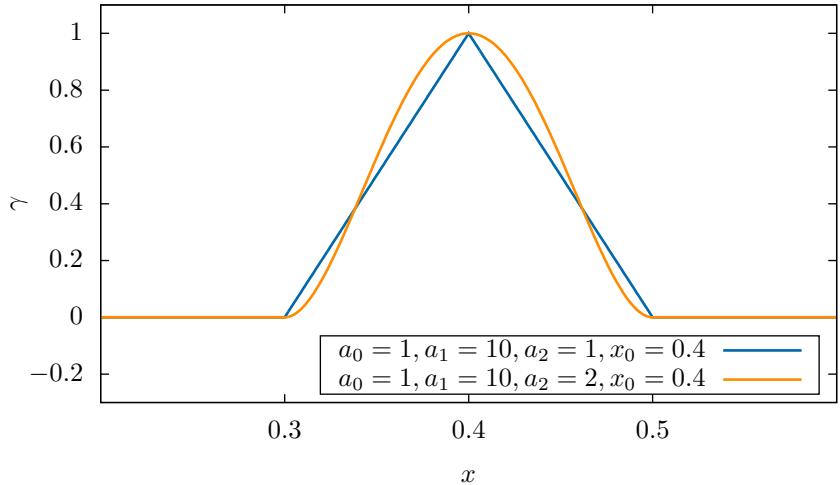}}
  \caption{Sketch of the test function $\gamma$ defined in~\eqref{eq:filter-function} in one dimension, for two values of the
    smoothing parameter $a_2$.}
  \label{fig:gamma}
\end{figure}

Finally, the test function $\gamma$ in the definition of $Y_{\gamma}^{\L}$ 
in~\eqref{eq:def_Y-N} is chosen as
\begin{equation}
  \label{eq:filter-function}
  x\mapsto\gamma(x)
  =a_0~\prod_{j=1}^d\left(\max\left(0,1-\left(a_1\abs{x^{(j)}-x_0^{(j)}}\right)^{a_2}\right)\right)^{a_2},
\end{equation}
with independent parameters $a_0, a_1>0$, $a_2\geq1$, and where $x^{(j)}$ denotes the $j^{\text{th}}$ Cartesian component of the
vector $x$. The graph of this function resembles a smoothed hat function centred at $x_{0}\in\Omega$ and symmetric with respect to
that point (see Figure~\ref{fig:gamma}), where the height is given by $a_0$, the support is of length $\frac{2}{a_1}$ in each
dimension; $a_2$ scales the smoothness: $a_2=1$ would correspond to a piecewise linear wedge-shaped graph in one dimension, while
$a_2=2$ gives a smoother function.  The dependence of the measured transport coefficients on these parameters will be discussed in
Subsection~\ref{sec:Parameter-dependence}. In practice, the same simulation data is post-processed with test functions
concentrated at multiple points $x_0$. This allows to obtain simultaneously the value of the transport coefficients at different
densities in the case of non-equilibrium evolutions or to increase the efficiency in measurements gathered from equilibrium data.

We note that we are using the same realisations for finding the deterministic states, $\rho(t_{0},\cdot)$ and
$\rho(t_{0}+h,\cdot)$, as well as for the expectation of the quadratic variation on the left-hand side
of~\eqref{eq:formula-Y-sigma}. This can lead to an underestimation of $\ip{(\nabla\gamma)^{2}}{\m(\rho(t_0,\cdot))}$. Thanks to
the choice of $F$ in~\eqref{eq:choice-F} as a quadratic function, this error can be compensated the same way as for the sample
standard deviation, i.e., by multiplying the left-hand-side of~\eqref{eq:formula-Y-sigma} with a factor of
$\frac{R}{R-1}$~\cite[Chapter 2]{Dixit2016a}. The method is summarised in Algorithm~\ref{alg:pseudocode}.

The actual measurements based on~\eqref{Eq:m_approx} give the quantity $\m$. To obtain the diffusivity $D$, we
use~\eqref{eq:d-and-m} in combination with the analytical expressions for the entropy $\S$ given in
Section~\ref{sec:Part-models-stud} for each of the three examples studied. We remark that the use of the analytic form of the
entropy is only for simplicity; in principle it could also be approximated from the particle data.

In addition to the measurement of the transport data $m$ and $D$, we also estimate the error bars of such measurement based on the
standard error of the expectation, associated to the $R$ samples. To focus on the key approximation errors, we neglect the error
arising from evaluating the deterministic state $\rho$ via the law of large numbers from the same realisations. We further ignore
this source of error when converting between $\m$ and $D$.  In principle, this leads to an underestimation of the statistical
error.

The particle processes described in the next section are modelled by a Lattice Kinetic Monte Carlo approach (L-KMC), and are
evolved according to the Bortz-Kalos-Le\-bo\-witz algorithm \cite{Bortz1975a}. Both the process and the proposed method to compute
the diffusivities are implemented in C.

\begin{algorithm}
\raggedright
\everypar={\nl}
\SetAlgoVlined

\tcp{Set lattice domain and scaling:}
$\delta x = \frac{1}{\L}$, $x_{i}=\frac{X_{i}}{\L}$			\tcp*{Spatial discretisation; $X_i$ is the lattice coordinate}

\tcp{Generate particle data from initial profile $\eta\left(t_{\ini}\L^2,X_{i}\right)$}:
\For{all $R_{1}$ realisations starting from $t_{\ini}$}
    {
      $\eta_{r}\left(t_{\prep}\L^2,X_{i}\right) = \text{stochastic-evolution}\left(\left[ t_{\ini}, t_{\prep}\right]
      ,\eta\left(t_{\ini}\L^2,X_{i}\right)\right)$ ;
      
      \For{all $R_{2}$ realisations starting from $t_{\prep}$}
          {
            $\eta_{r}\left(t_{0}\L^2,X_{i}\right) = \text{stochastic-evolution}\left(\left[ t_{\prep}, t_{0}\right],
            \eta_{r}\left(t_{\prep}\L^2,X_{i}\right)\right)$ ;
            
            $\eta_{r}\left(\left(t_{0}+h\right)\L^2,X_{i}\right) = \text{stochastic-evolution}
            \left(\left[ t_{0}, t_{0}+h\right],\eta_{r}\left(t_{0}\L^2,X_{i}\right)\right)$ ;
          }
    }
    $R = R_{1}\cdot R_{2}$ 		\tcp*{Total number of realisations in $\left[t_{0},t_{0}+h\right]$}
    
    $\rho\left(t_{0},x_{i}\right)
    = \frac{1}{R}\cdot \sum_{r}\left(\eta_{r}\left(t_{0}\L^2,X_{i}\right)\right)$ 	\tcp*{Approx.~deterministic state}

    $\rho\left(t_{0}+h,x_{i}\right)
    = \frac{1}{R}\cdot \sum_{r}\left(\eta_{r}\left(\left(t_{0}+h\right)\L^2,X_{i}\right)\right)$ ;

    \tcp{Test function $\gamma$ with parameters $a_0$, $a_1$, $a_2$, $x_0$:}
    
    \SetKwProg{Fn}{Function}{}{}
    
    \Fn{$\gamma\left(x_{i}\right)$}
       {
         $\gamma\left(x_{i}\right) =  a_0  \prod_{j=1}^d
         \left(\max\left(0,1-\left(a_1\abs{ x_{i}^{\left(j\right)} -x_{0}^{\left(j\right)}}\right)^{a_2}\right)\right)^{a_2}$ ;

         	\tcp*{$j$ is spatial index} 
         \KwRet{$\gamma\left(x_{i}\right)$} ;
       }
       
       \tcp{Compute denominator of Equation~\eqref{Eq:m_approx}:} 
         
         \For{all lattice positions $X_{i}\in \Lambda$}
             {
               
               $G_{i} = \sum_{j=1}^d\left( \frac{\gamma\left(x_{i}+\delta x\, e^{\left(j\right)}\right)
                 - \gamma\left(x_{i}-\delta x\, e^{\left(j\right)} \right)}{2 \delta x} \right)^{2}$ 	
 \tcp*{$e^{\left(j\right)}$ is $j$-th Cartesian unit vector} 
             }
             $G = 2 \left(\delta x\right)^{d}~ \sum_{i}\left(G_{i}\right)$
             \tcp*{$2\norm{\nabla \gamma}_{L^{2}}^{2}$}

             \tcp{Compute numerator of Equation~\eqref{Eq:m_approx}:}

             \For{all $R$ realisations $r$}
                 {
                   
                   \For{all microscopic positions $X_{i}\in \Lambda$}
                       {
                         $Y_{r}\left(t_{0},x_{i}\right) = \sqrt{\L^{d}} 
                         \left( \eta_{r}\left(t_{0} \L^{2},X_{i}\right) -\rho\left(t_{0},x_{i}\right)\right)$ ;

                         $Y_{r}\left(t_{0}+h,x_{i}\right) = \sqrt{\L^{d}} 
                         \left(\eta_{r}\left(\left(t_{0}+h\right) \L^{2},X_{i}\right) -\rho\left(t_{0}+h,x_{i}\right)\right)$ ;
                       }
                       $M_{r} = (\delta x)^d \sum_{i}\left( \gamma \left(x_{i}\right) \left( Y_{r}\left(t_{0}+h,x_{i}\right)
                       - Y_{r}\left(t_{0},x_{i}\right) \right)\right)$ ;
                 }
                 $M =  \frac{1}{h} \frac{1}{R-1}  \sum_{r}\left(M_{r}^{2}\right)$
                 \tcp*{$\frac{1}{h} \mathbb{E}\left[\left[Y_\gamma(t_0+h)-Y_\gamma(t_0)\right]^{2}\right]$}
                 
                 \tcp{Result:}
                 $\m = \frac{M}{G}$ 	\tcp*{Mobility}

                 \caption{Pseudo-code describing the method in algorithmic form.}
                 \label{alg:pseudocode}
\end{algorithm}

\section{Particle models studied}
\label{sec:Part-models-stud}

In this section, we describe the two types of particle processes studied here.  These processes have been chosen since they allow
for a comparison with analytic solutions, for the chosen parameters.  They are Markovian, composed of indistinguishable particles
and have a hydrodynamic limit of the form~\eqref{eq:SPDE-det}. Specifically, we consider two zero range processes (one of them
being the special case of Brownian particles) and a simple exclusion process.

\subsection{Zero range process}
\label{sec:Zero-range-process}

\emph{Zero range processes (ZRPs)} are particle processes on a lattice, where finitely many particles (possibly none) are located
on each lattice site.  The jump rate at which one particle leaves a site $X$ depends only on the total number $k(X)$ of particles
at this site $X$ and it is described by the (local) \emph{jump rate function} $g\colon\N_0\to\R^+_0$. We consider two cases:
$g(k) = k$, which corresponds to independent Brownian particles (i.e., linear diffusion equation as its hydrodynamic limit), and
$g(k) = k^2$, which also has an analytic expression for the diffusivity.

The process starts at some initial configuration $\eta$. The system waits an exponential microscopic time drawn from a Poisson
distribution with rate $\lambda(\eta):=\sum_{X\in\Lambda}g(\eta(X))$, at which time one particle is moved from $X$ to $\tilde X$
with probability $\frac{g(\eta(X))}{\lambda(\eta)}p(\tilde X-X)$; here we choose for simplicity
\begin{equation}
  p(\tilde X -X) = 
  \begin{cases}
  \frac{1}{2^d} & \text{if }\left| \tilde X - X\right|=1\\
  0 & \text{otherwise}
  \end{cases}; 
\end{equation}
so particles jump to their nearest neighbour only, with equal probability. After this jump, the process starts again from
$\eta^{X,\tilde X}$, which is the configuration where one particle has changed its position from $X$ to $\tilde X$.

The hydrodynamic limit of ZRPs is in general a nonlinear diffusion equation~\eqref{eq:SPDE-det}. Namely, let $\rho^\L$ denote the
diffusively rescaled density representing the particle process. That is, for positions $x=X/L$ with $X\in\Lambda$, we set
$\rho^\L(t,x) := \eta(t \L^2, x \L)$ 
and interpolate in a piecewise constant manner in between. Then formally, in a suitable weak sense,
\begin{equation*}
  \rho^\L(t,x) \to \rho(t,x) 
\end{equation*}
and $\rho$ solves~\eqref{eq:SPDE-det} (for the precise formulation in a measure setting see~\cite{Kipnis1999a}).

The \emph{thermodynamic entropy of the zero range process}, in dimensionless units, is~\cite{Grosskinsky2003a}
\begin{equation}
  \label{eq:thermo-ent}
  \S(\rho) = \int_\Omega \left[-\rho \log (2\m(\rho)) + \log Z(2\m(\rho)) \right]\df x, 
\end{equation}
with the \emph{partition function} $Z$
defined as $Z(\varphi):=\sum_{k\in\N_0}\frac{\varphi^k}{g!(k)}$; here $g!(k):=g(1)\cdot g(2) \cdot\ldots\cdot g(k)$.

In general, this formula does not lead to explicit expressions for $\m$. Yet, here we consider two cases for which analytic
expressions are available.  The first case considered is the linear one, $g(k)=k$, which corresponds to Brownian particles, and
for which $\m(\rho)\equiv \frac{1}{2}\rho$. The second case is $g(k) = k^2$, where $\m(\rho)$ is implicitly given as the inverse
of $\rho\left(\frac{\m}{2}\right)=\sqrt{\m}\cdot\frac{I_{1}(2\sqrt{\m})}{I_{0}(2\sqrt{m})}$, with $I_{i}$ being the modified
Bessel-function of the first kind. This can be derived directly from the definition of these Bessel functions in combination with
the equilibrium measure
$\bar{\nu}_{\frac{\m}{2}} \left(\eta\left(T,X_{i}\right)=k\right)=\frac{1}{Z\left(\m\right)}\cdot\frac{\m^{k}}{g!\left(k\right)}$
and $\rho\left(\m\right)=\mathbb{E}_{\bar{\nu}_{\m}}\left[\eta\left(T,X_{i}\right)\right]$. Indeed,
\begin{multline*}     
  \rho\left(\frac{\m}{2}\right)=\mathbb{E}_{\bar{\nu}_{\frac{\m}{2}}}\left[\eta(T,X_{i})\right] =
   \left.
   \frac{\sum_{k=0}^\infty  k\frac{\m^{k}}{g!(k)}}{\sum_{k=0}^\infty\frac{\m^{k}}{g!(k)}} 
  \right|_{g!(k)=(k!)^{2}} \\
   =\frac{ \sum_{k=1}^\infty\frac{\left(\frac{2\sqrt{\m}}{2}\right)^{2k}}{(k-1)!\cdot k!}}
    {\sum_{k=0}^\infty\frac{\left(\frac{2\sqrt{\m}}{2}\right)^{2k}}{k!\cdot k!}} =
   \frac{ \sqrt{\m}\cdot\frac{2\sqrt{\m}}{2}
   \sum_{k=1}^\infty\frac{\left(\frac{2\sqrt{\m}}{2}\right)^{2(k-1)}}{(k-1)!\cdot k!}}
    {\sum_{k=0}^\infty\frac{\left(\frac{2\sqrt{\m}}{2}\right)^{2k}}{k!\cdot k!}} 
     =\sqrt{\m}\frac{I_{1}\left(2\sqrt{\m}\right)}{I_{0}\left(2\sqrt{\m}\right)}, 
  \end{multline*}
  where the last equality may be found in~\cite[(9.6.10)]{Abramowitz1964a}. It then follows with a short calculation
  from~\eqref{eq:thermo-ent} and~\eqref{eq:SPDE-det} that the hydrodynamic limit can be written as a nonlinear diffusion equation
  of the form
\begin{equation*}
  \partial_t\rho = \K(\rho) \DS(\rho) =-\div(\m(\rho) \nabla \DS(\rho)) = \Delta(\m(\rho)). 
\end{equation*}
For the zero range process, $\frac{1}{2}\Phi(\rho) = \m(\rho)$~\cite[Chapter 5, Theorem 1.1]{Kipnis1999a}. Thus, the jump rate $g$
determines $\m$ (normally implicitly, with two explicit examples given above); $\m$ and $\Phi$ are identical up to a
prefactor. This defines all parameters in~\eqref{eq:fluctuation} and~\eqref{eq:def_Y} for the zero range process.

\subsection{Simple exclusion process}
\label{sec:Simple-excl-proc}

In the \emph{simple exclusion process} (SEP), particles attempt to jump to neighbouring sites with a constant rate one. However,
if the destination site is already occupied, the jump is abandoned and the particle stays at its current location. Consequently,
all sites are occupied by at most one particle. The jump rate from site $X$ to neighbouring site $\tilde X$ is
$g_{X\rightarrow \tilde X}(\eta):=\frac{1}{2^d}\eta(X)(1-\eta(\tilde X))$. For further details,
see~\cite[Section~2.2]{Kipnis1999a}.  For the simple exclusion process, $\m(\rho)= \frac{1}{2}\rho(1-\rho)$~\cite{Adams2013a} and
$\Phi(\rho) = \rho$~\cite[Chapter 4, Theorem 2.1]{Kipnis1999a}.  We remark, however, that in the one-dimensional setting studied
numerically in Section~\ref{sec:Comp-results} the individual particles themselves are not following a Brownian motion on a
microscopical level~\cite{Arratia1983a}. The entropy of the simple exclusion process is the \emph{mixing entropy}, i.e.,
$\S(\rho) = - \int\left[ \rho \log \rho + (1-\rho) \log(1-\rho)\right] \df x$ in dimensionless form. This defines all quantities
in~\eqref{eq:fluctuation} and~\eqref{eq:def_Y} for the simple exclusion process.

\section{Computational results}
\label{sec:Comp-results}

We show simulation results in one space dimension for each of the three processes discussed in Section~\ref{sec:Part-models-stud},
namely two zero range processes, one with $g(k) = k$, i.e., random walkers, and one with $g(k) = k^2$, and a simple exclusion
process. All results shown are given for non-equilibrium situations, i.e., starting from non-constant initial profiles
$\eta(t_{\ini}L^2,X)$. This allows us to obtain diffusivity information for a wide range of densities within a single set of
simulations, by choosing different concentration points $x_{0}$ for the test functions $\gamma$. Note, however, that, if these
$x_{0}$ are chosen too close to each other, their results might not be statistically independent anymore. This can be addressed,
by monitoring the correlations and suitable post-processing.

\subsection{Default choice of parameters}
\label{sec:Defa-choice-parame}

For better comparability, the following default settings are used, unless stated otherwise. The initial profile is taken as
$\eta(t_{\ini}L^2,X) =25 \sin\left(\pi \frac{X}{\L}\right)$ for the zero range processes, and
$\eta(t_{\ini}L^2,X) =0.95 \sin\left(\pi \frac{X}{\L}\right)$ for the simple exclusion process. The lattice length is $\L=5\,000$,
and the chosen time parameters are such that $t_{\prep}-t_{\ini}=4\cdot10^{-6}$ and $t_0 - t_{\prep} = 4\cdot10^{-9}$, and the
measurement time is $h=4\cdot10^{-11}$. We take $R_1 = 50$ and $R_2 = 2\,000$, so in total $R = 100\,000$ realisations are
simulated. The parameters chosen for the test functions are $a_0=1$, $a_1=160$, $a_2=2$. Further, $39$ points uniformly
distributed over the unit interval are chosen as concentration points $x_0$. Furthermore, given the symmetry of both the chosen
profile and the concentration points for the test functions $\gamma$, we will average results for similar densities obtained from
the left and right half of the $\sin$ profile, to make the plots more readable.

\begin{figure} [htbp]
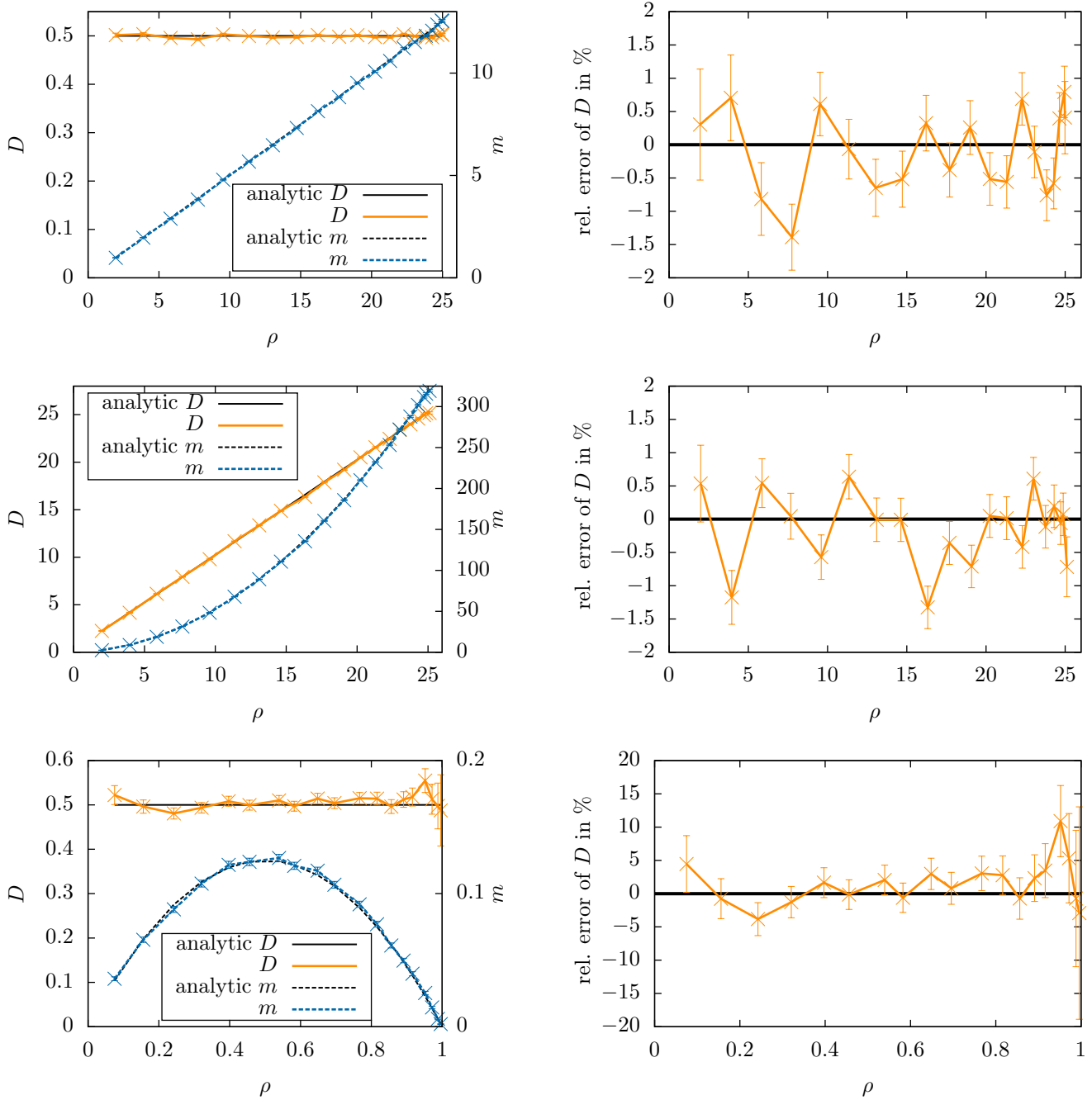

\centering
  \centerline{\input{"ZRP_rw_sin-rho-comparison_Celia.tex"} \input{"ZRP_rw_sin-rho-comparison_relerr_Celia.tex"}}
  \centerline{\input{"ZRP_rep_sin-rho-comparison_Celia.tex"} \input{"ZRP_rep_sin-rho-comparison_relerr_Celia.tex"}}
  \centerline{\input{"SEP_sin-D_m_vs_rho-comparison.tex"} \input{"SEP_sin-rho-comparison_relerr.tex"}} 
  \caption{Measurements of the transport coefficients from non-equilibrium evolutions corresponding to a random walk (top row), a
    ZRP with $g(k) = k^2$ (middle row) and a SEP (bottom row).  Shown are the diffusivity $D$ and the mobility $\m$ (left panel)
    and the relative errors as compared to the analytical results (right panel). The choice of parameters is the default one,
    described in Subsection~\ref{sec:Defa-choice-parame}.  }
  \label{fig:d-and-m}
\end{figure}

\subsection{Results and comparison to the analytical solutions}
\label{sec:results}

Figure~\ref{fig:d-and-m} displays, for each process, the diffusivity $D$ and mobility $\m$, where the former is computed from the
latter via~\eqref{eq:d-and-m} (see Section~\ref{sec:Part-models-stud} for the expressions of the entropy for each process), i.e.,
\begin{equation}
  \label{eq:D_from_m}
  D(\rho) = 
  \begin{cases}
    \dfrac{m\left(\rho\right)}{\rho} & \text{for the ZRP with }g\left(k\right)=k ,\\
    \dfrac{m\left(\rho\right)}{\Phi\left(\rho\right)-\rho^{2}} & \text{for the ZRP with }g\left(k\right)=k^{2},\\
    \dfrac{m\left(\rho\right)}{\rho \left(1-\rho\right)} & \text{for the SEP}.
  \end{cases}
\end{equation}
To better assess the accuracy of the method, the right panels show the relative error of $D$, i.e., $(D-D_{\ana})/{D_{\ana}}$,
which coincides with that of $m$.  We find the agreement between simulation and analytic solution very strong: the relative errors
for the zero range processes are mostly below the single percentage range. Relative errors for the simple exclusion process are
about one order of magnitude larger for the same choice of the parameters. This is to be expected, as an untypical behaviour of a
single particle can block other particles for long times. Thus, simulations of the simple exclusion process require finer
discretisation in order to obtain the same accuracy as for the zero range process. Yet, for better comparison, the parameters have
been chosen uniformly for all the processes.

\subsection{Parameter dependence}
\label{sec:Parameter-dependence}

\begin{figure} [htbp]
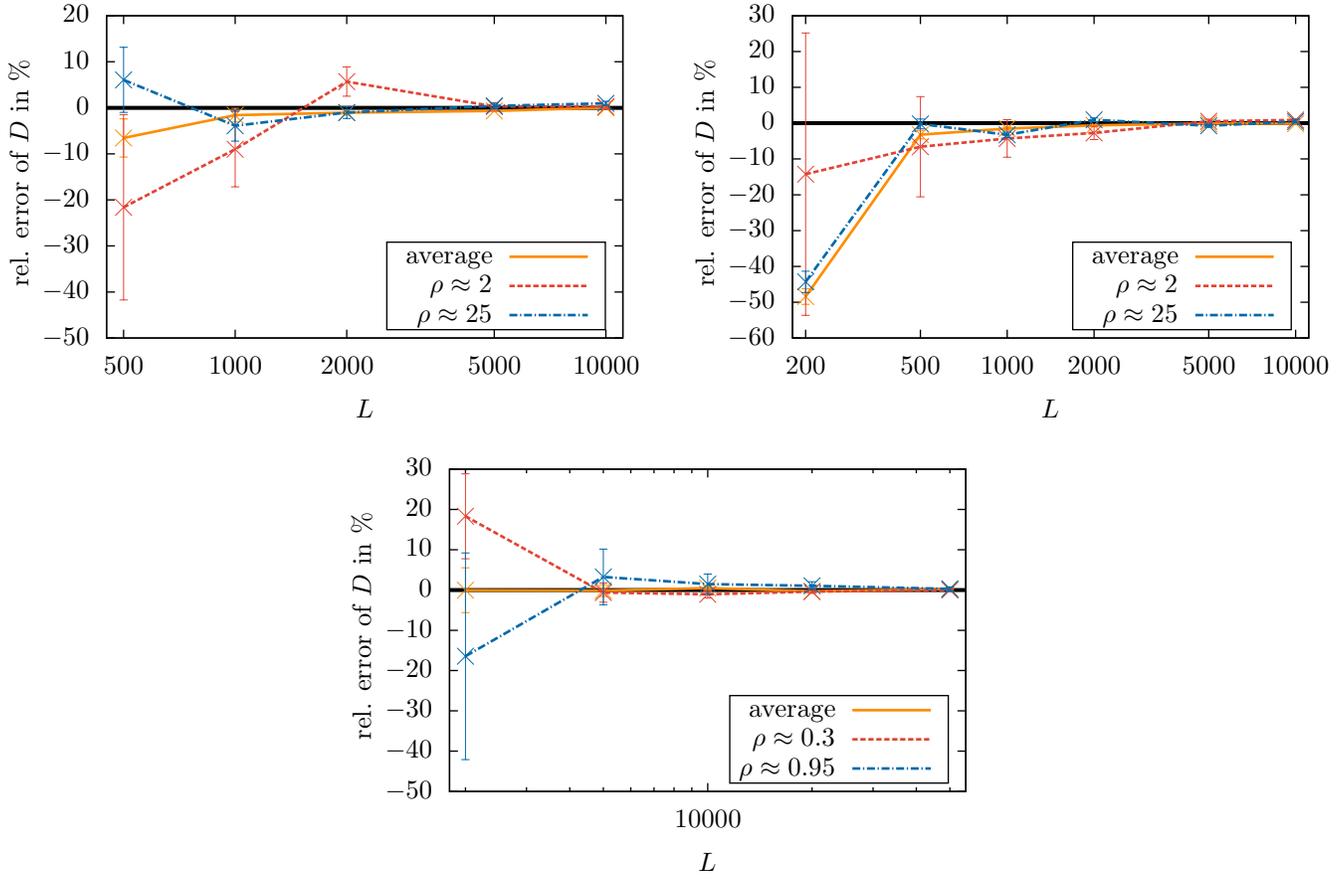

\centerline{\input{"ZRP_rw_sin-Nbin-comparison_relerr_Celia.tex"} \input{"ZRP_rep_sin-Nbin-comparison_relerr_Celia.tex"}}
\centerline{\input{"SEP_sin-Nbin-comparison_relerr.tex"}}
\caption{Relative errors in the diffusivity calculation with respect to the sample size $\L$ for the ZRP with $g(k)=k$ (top left),
  the ZRP with $g(k)=k^2$ (top right) and the SEP (bottom). For the parameters and non-equilibrium initial profile we refer to
  Subsection~\ref{sec:Defa-choice-parame}.}
  \label{fig:details_L}
\end{figure}

\begin{figure} [htbp]
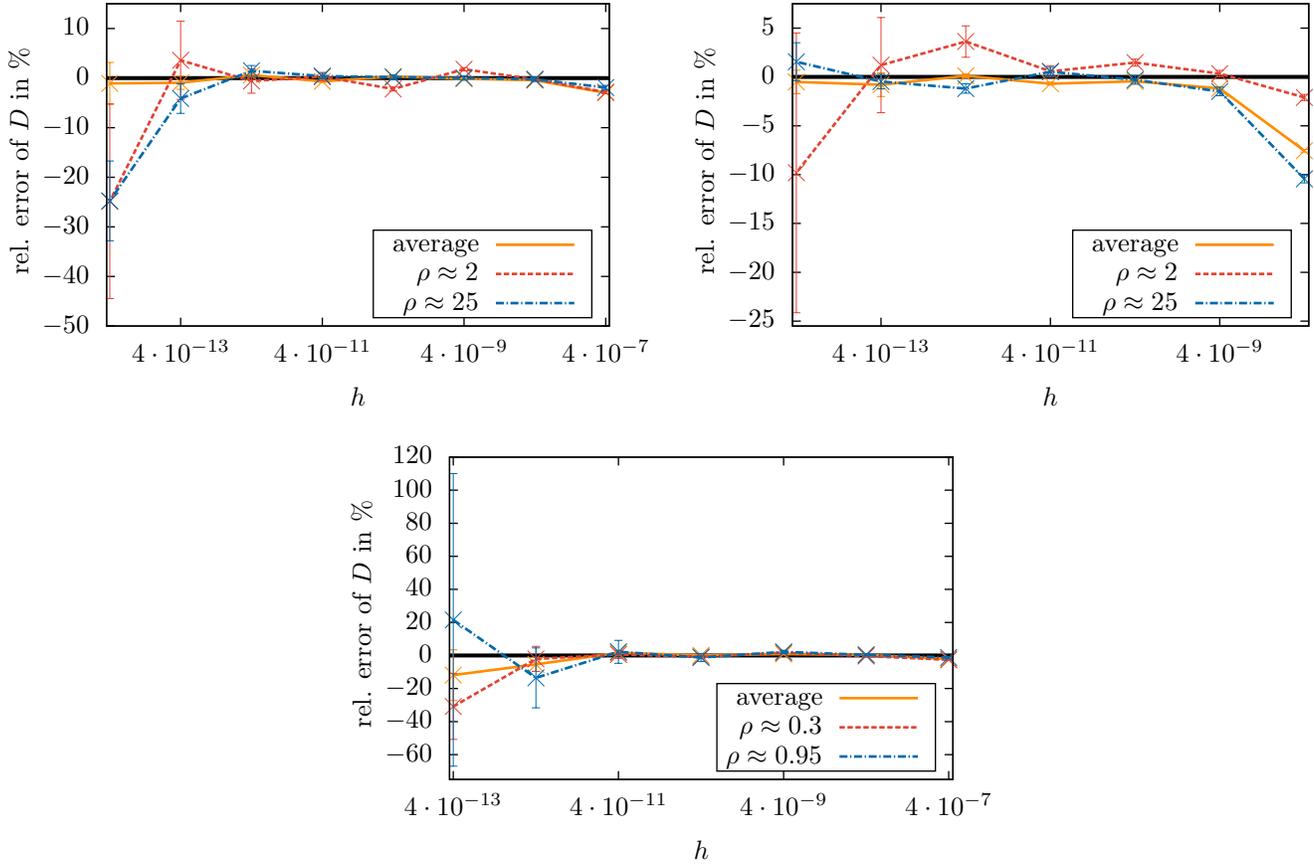

  \centerline{\input{"ZRP_rw_sin-dT-comparison_relerr_Celia.tex"} \input{"ZRP_rep_sin-dT-comparison_relerr_Celia.tex"}}
  \centerline{\input{"SEP_sin-dT-comparison_relerr.tex"}}
  \caption{Relative errors in the diffusivity calculation with respect to the measurement time $h$ for the ZRP with $g(k)=k$ (top
    left), the ZRP with $g(k)=k^2$ (top right) and the SEP (bottom). The parameters are the default ones described in
    Subsection~\ref{sec:Defa-choice-parame}.}
  \label{fig:details_h}
\end{figure}

\begin{figure} [htbp]
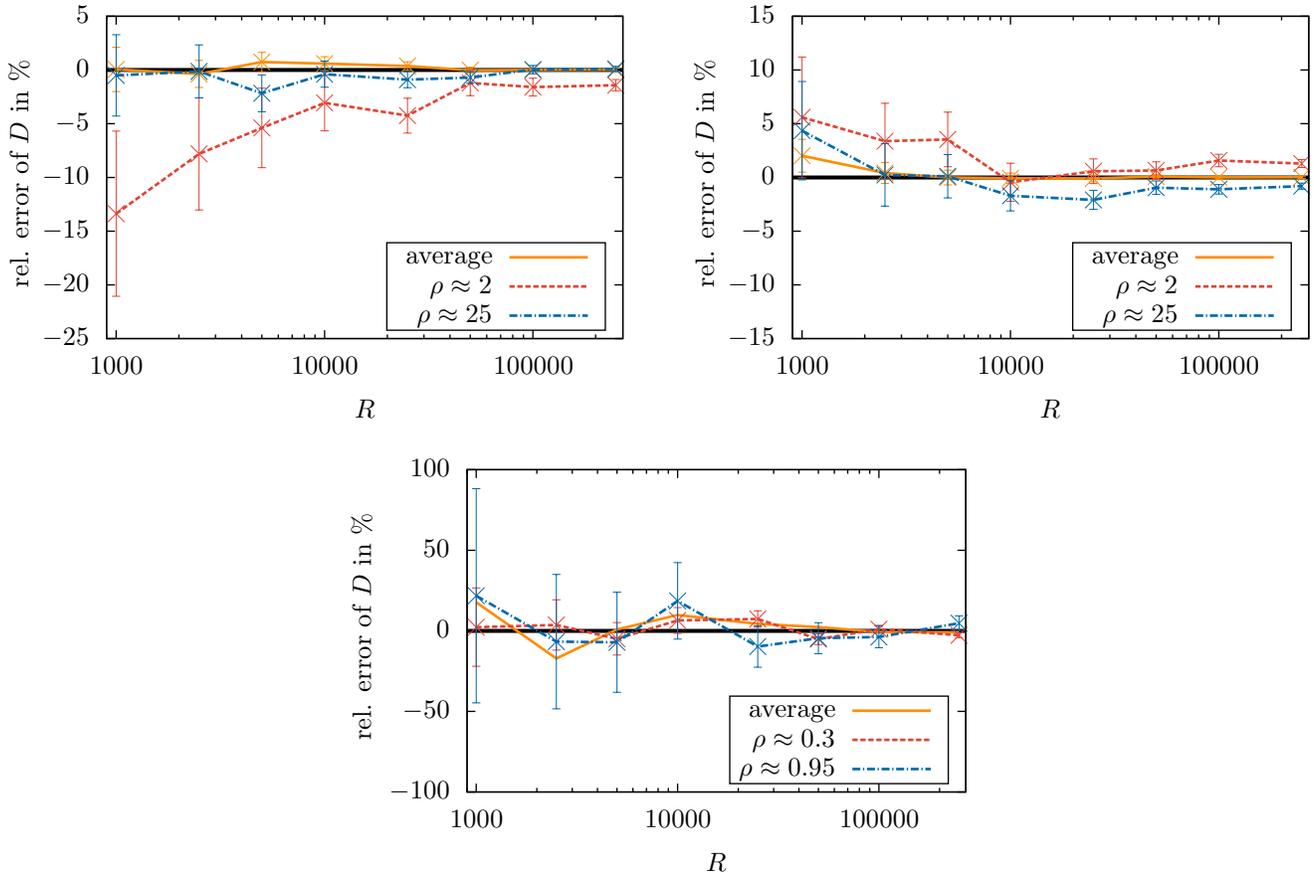

  \centerline{\input{"ZRP_rw_sin-overallsampling-comparison_relerr_Celia.tex"}
    \input{"ZRP_rep_sin-overallsampling-comparison_relerr_Celia.tex"}}
  \centerline{\input{"SEP_sin-overallsampling-comparison_relerr.tex"}}
  \caption{Relative errors in the diffusivity calculation with respect to the sample size $R$ for ZRP with $g(k)=k$ (top left),
    ZRP with $g(k)=k^2$ (top right) and SEP (bottom). Here $R_{1}=50$ and $R_{2}$ was varied to achieve the different values of $R
    = R_{1}\cdot R_{2}$. For all other parameters, the default settings of Subsection~\ref{sec:Defa-choice-parame} apply.}
  \label{fig:details_R}
\end{figure}

In Figures~\ref{fig:details_L} to~\ref{fig:details_R} we show the dependence of the results on the choice of the parameters $\L$,
$h$ and $R$ for all three processes. In each figure, we depict the errors associated with a representative low and high value of
the density, as well as the average error over the multiple densities considered (namely, over the results of the different
concentration points $x_0 \in \{0.1,0.2,\dots,0.9\}$ chosen for the test function $\gamma$). Regarding $L$, the proposed numerical
strategy relies on having a large enough system size, as only then we can expect the evolution of fluctuations $Y^{\L}$ to be well
approximated by an Ornstein-Uhlenbeck-process~\eqref{eq:fluctuation}.  This convergence is depicted in Figure~\ref{fig:details_L},
where the errors are shown to decrease with increasing values of $L$, as anticipated.  Also, the method requires the measurement
time $h$ to be short enough, so that the limit on the left-hand-side of~\eqref{eq:formula-Y-sigma} is well approximated. For the
simple exclusion process, systematic deviations become visible just at around $h\approx4\cdot10^{-7}$, whereas for the zero range
process with $g(k)=k^2$ aberrations already start at $h\approx4\cdot10^{-9}$. We remark that for too small values of $h$,
artificial errors are also to be expected, since too few particles might jump and the system might not yet exhibit its diffusive
behaviour, i.e., the limit $h\to0$ might not commute with the limits $R\to\infty$ or ${\L\to\infty}$. Finally, good estimates of
the expectation (both the left-hand-side of~\eqref{eq:formula-Y-sigma} and the deterministic states $\rho$) require by the Law of
Large Numbers large sample sizes $R$. This convergence is depicted in Figure~\ref{fig:details_R}, where both the errors and the
error bars tend to zero with increasing values of $R$. We note that the error bars exhibit an almost perfect power law behaviour
of the form $c R^{-1/2}$ with some constant $c$, as expected from the Central Limit Theorem. No clear power-law behaviour is
observed for the errors themselves as a function of $R$.

\begin{figure}
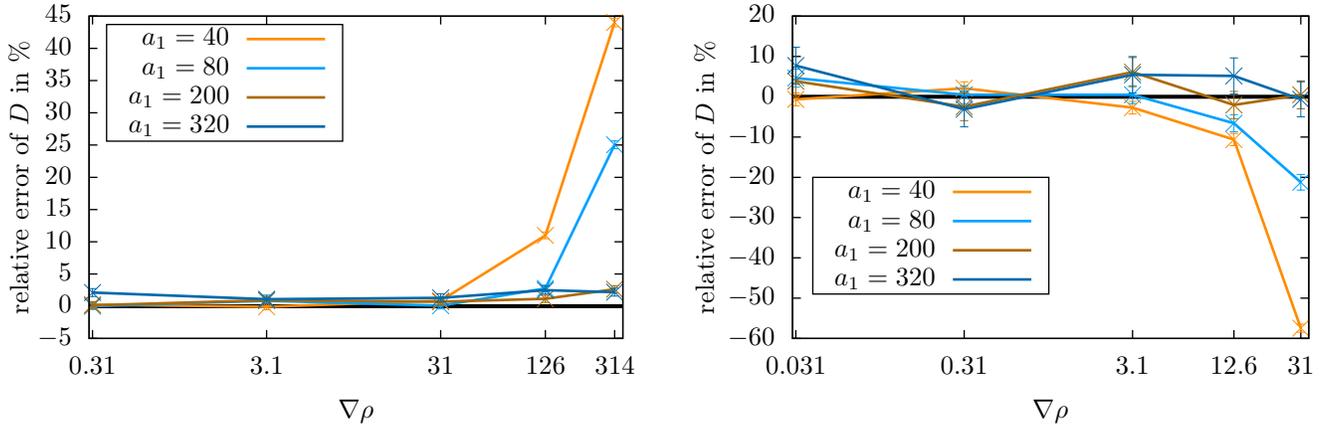

  \centerline{\input{"ZRP_rep_slope-comparison.tex"} \input{"SEP_slope-comparison.tex"}}
  \caption{Relative errors in the diffusivity calculation with respect to the slope of the density profile $\nabla \rho$ and the
    test function parameter $a_1$, for the ZRP with $g(k) = k^2$ (left) and the SEP (right). The non-equilibrium profiles
    considered are described in Subsection~\ref{sec:Parameter-dependence}, and all other parameters are set to their default
    values of Subsection~\ref{sec:Defa-choice-parame}.}
  \label{fig:N-dependence_b-dependence}
\end{figure}

We further study in Figure~\ref{fig:N-dependence_b-dependence} the dependence of the measurement errors on the initial profile
$\eta(t_{\ini}L^2,X)$ and the parameters of the test function $\gamma$ of~\eqref{eq:filter-function}. Of particular interest is
the interplay of the parameter $a_{1}$, which is inversely related to the support of the test function $\gamma$, and the local
slope $\nabla \rho$ of the initial profile at the point of measurement, which serves to quantify how far the system is from
equilibrium.  To demonstrate this, we consider various initial profiles so as to measure the transport coefficient at a point of
constant density and varying slope. More specifically, we consider
$\eta\left({t_\ini}L^2,X\right) = 5 + 5 \sin \left( \frac {2 \pi}{A} \left( \frac{X}{\L} - \frac{1}{2} \right) \right)$ for case
of the ZRPs and
$\eta\left({t_\ini}L^2,X\right) = \frac{1}{2} + \frac{1}{2} \sin \left( \frac {2 \pi}{A} \left( \frac{X}{\L} - \frac{1}{2} \right)
\right)$
for the SEP, where $A\neq0$ is the parameter that controls the slope at the chosen point of measurement, here
$x_{0} = \frac{1}{2}$.  As shown in the figure, being far from equilibrium can have an impact on the outcome, where the precise
value of the error depends on the process.  In all cases, however, this (unwanted) dependence can be cured by choosing a more
narrowly supported test-function (i.e., larger $a_{1}$). We note that for a fixed finite system size $\L$, the support of the test
function $\gamma$ can only be narrowed until a limit determined by the lattice spacing is reached, unless the expression
$\ip{\nabla \gamma}{\nabla \gamma}$ is evaluated analytically.

Regarding the other two parameters of the test function $\gamma$, we remark that the derivation of the algorithm assumed
$\gamma \in C^2_0$, which is satisfied for the choice $a_2 >2$. Yet, simulations for $a_2 = 2$ or even $a_2=1$ still delivered
good results, with slightly reduced accuracy for $a_2=1$. Finally, $a_0$ does not play any role, as it drops out algebraically
in~\eqref{eq:formula-Y-sigma}.

\subsection{Alternative method for sequence of consecutive measurements}
\label{sec:Altern-meth-sequ}

\begin{figure}
  \centerline{\input{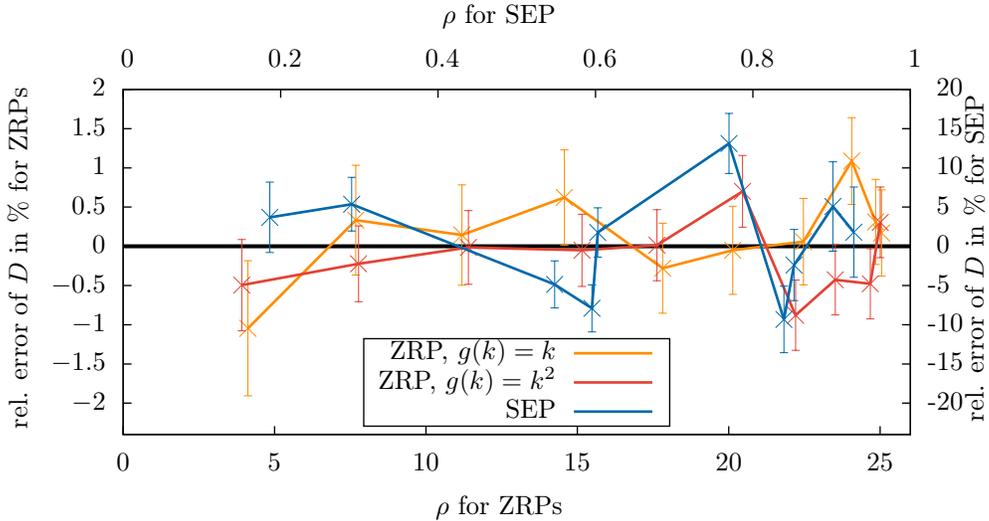}}
  \caption{Results for the diffusivity $D$ for all three processes (a random walk, a ZRP with $g(k) = k^2$ and a SEP), based on
    one sequence of snapshot measurements as described in Subsection~\ref{sec:Altern-meth-sequ}.  Here $P =100\,000$ measurements
    were made and for all other parameters the default choices in Subsection~\ref{sec:Defa-choice-parame} are used.}
  \label{fig:continuous_m}
\end{figure}

One drawback of the method described above is that one must be able to prepare the same initial state $\eta\left(t_0L^2, X\right)$
numerous times to obtain a good estimate of the deterministic $\rho$ in~\eqref{eq:formula-Y-sigma}. This may be difficult in some
experimental settings, where it could be advantageous to observe the system over only one, longer, period of time.  In such a
scenario, one would take measurements at snapshots, say at times $h, 2h, 3h,\dots, P\cdot h$ with $P\in\N$, instead of analysing
multiple realisations at times $t_0$ and $t_0+h$.  We sketch here how this alternative strategy might be adopted in the
experimental setting described.  Towards this purpose, we deviate from the approach described in Subsection~\ref{sec:Comp-strat}
by substituting $Y^{\L}_{\gamma}$ from~\eqref{eq:def_Y-N} with
\begin{equation*}
  \tilde{Y}^\L_\gamma(t) := \sqrt{\L^d} \ip{\gamma}{\rho^\L(t,\cdot)}. 
\end{equation*}
That is, we suppress the deterministic states entirely. Note that $Y^{\L}_{\gamma}$ only appears in the function $F$
in~\eqref{eq:choice-F}. To use $\tilde{Y}^\L_\gamma(t)$, one would require the difference $\rho\left(t_{0}+h\right) -
\rho\left(t_{0}\right)$ to be negligible for $h\to 0$, when compared to the difference of their stochastic
counterparts. Obviously, if the deterministic state is differentiable, then this difference is $\mathcal{O}(h)$, which should be
of lower order. Yet, it is not clear that this is preserved in the second limit $\L \rightarrow \infty$. This makes the version
sketched in this subsection more speculative, and a detailed investigation will be the topic of a separate investigation. Yet,
initial computational results indicate a good agreement of the original approach of Subsection~\ref{sec:Comp-strat} and the
modification discussed in this subsection. We call the former method \emph{parallel} and the latter \emph{sequential}.  When
measuring in equilibrium, the sequential and the parallel method gave equivalent results. Figure~\ref{fig:continuous_m} shows the
diffusivity $D$ computed with the sequential method for non-equilibrium data. The relative errors are found to be in good
agreement with those of the parallel method.

\subsection{Discussion}
\label{sec:Discussion}

While the simulations deliberately had a sinusoidal initial profile to probe the method out of equilibrium, the errors on the
diffusivities are even lower for constant initial profiles.  In such a setting, equilibrium methods are applicable, such as mean
square displacement (MSD). We here compare the performance of both approaches (MSD and the parallel method proposed in this
article) for the zero range process with $g(k)=k$ (the random walk) and $g(k)=k^2$. {We remark that in }
the standard mean square displacement method, particles are tagged and the diffusivity is computed from the slope of the temporal
evolution of the mean square displacement. The standard setting assumes that the evolution of a tagged particles scales as
$X(t) \sim \sqrt{t}$. Since the SEP in one space dimension scales as $X(t) \sim t^{1/4}$ and is thus inaccessible for this method,
the results for the method we propose cannot be compared. It is noteworthy that the method proposed here does compute the
diffusivity accurately also for this process.

To achieve a fair comparison between MSD and the proposed methodology, the algorithms of both methods have been implemented in
serial on the same machine, for systems of equal size $\L = 5\,000$, and with equal equilibration times. Also, the computation
time for the equilibration was ignored for the performance comparison, and we only generated new samples for our method with
$t_{0}-t_{\prep} = 4\cdot 10^{-9}$, i.e., we set $R_1 = 1$. Additionally, we assumed that the measurements can be repeated
independently arbitrarily often, leading to a scaling of the standard error of $m$ proportional to
$\nicefrac{1}{\sqrt{\text{computation time}}}$. For the same computation time, it turns out that the particle-tagging has a lower
error by a factor of about $17$ for the random walk and $2$ for the zero range process with $g(k)=k^2$, when compared to the
method proposed here. This effect is largely due to the relatively long preparation time $t_{0} - t_{\prep}$, as well as the extra
effort required for storing the measured states to compute the averages of both the deterministic states and the quadratic
variation. For $t_{0} - t_{\prep}=0$ (i.e., all initial states are microscopically identical), the factors reduce to $12$ and
$0.7$, respectively; i.e., the proposed parallel method is slightly faster than the established mean square displacement for the
ZRP with $g(k) = k^2$, but slower for the random walk. When using the sequential method proposed in
Subsection~\ref{sec:Altern-meth-sequ}, these factors reduce to $3$ and $0.2$, respectively. Note that the proposed approach is
trivially parallelisable, and taking advantage of this feature would dramatically accelerate the calculations.

We also emphasise that the method presented here applies to non-equilibrium evolutions, and that it enables the simultaneous
measurement of diffusivities over the range of densities present in the simulations (in contrast, equilibrium measurements deliver
the value of the transport coefficient at the single density simulated). Furthermore, we point out that our new approach does not
depend on any microscopic information, but only requires the macroscopic states. This makes it applicable in cases like the simple
exclusion process, where the individual particles are not exhibiting a diffusive behaviour, and could be particularly beneficial
in physical experiments or in social sciences, where microscopic data is less accessible. Finally, we note that the proposed
strategy only requires the initial and final data (at times $t_0$ and $t_0+h$), while other methods, such as Green-Kubo or mean
square displacement, require the full temporal evolution in the simulated time interval.

\section{Conclusions and outlook}
\label{sec:Conclusions-outlook}

This article considers macroscopically diffusive systems and presents a novel strategy for computing the diffusivity from
fluctuations in underlying stochastic particle systems. The method works in a wide range of out of equilibrium scenarios;
specifically, it only requires that the system is in local equilibrium and that it exhibits Gaussian fluctuations. As paradigm for
out of equilibrium evolution, sinusoidal initial profiles are simulated for particle models for which analytic expressions for the
diffusivity exist, and an excellent accuracy is observed. In addition, in equilibrium, the method compares well in terms of
computational cost with mean square displacement, when tested with the same processes, both in serial. The algorithm is trivially
parallelisable, and taking of advantage of this feature could offer dramatic speedup.

The method introduced in this article can in principle be extended to an even wider range of problems. The thermodynamic setting
of Section~\ref{sec:Therdm-metr-theory} generalises as follows to purely diffusive systems of the form
\begin{equation}
  \label{eq:entropy_gradient_flow}
  \partial_{t}\rho= \K(\rho) \DS(\rho),
\end{equation}
where $\K$ is a positive semidefinite operator and $\S$ is the entropy;~\eqref{eq:diff-linear-GF} is a special case of this
form. The extension of~\eqref{eq:dean} (or more generally~\eqref{eq:dean-strategy}) reads
\begin{equation*}
  \partial_t \rho^{\L} = \K(\rho^{\L}) \frac{\delta \S}{\delta \rho^{\L}} (\rho^{\L}) + \frac 1 {\sqrt {\L^d}} \sqrt{2
    \K(\rho^{\L})} \dot W_{x,t} ;
\end{equation*}
see~\cite[Eq.~(160)]{Eyink1990a}, or~\cite[Eqs.~(1.56)--(1.57)]{Ottinger2005a}. The fluctuation-dissipation relation establishes
that the fluctuation operator is directly linked to the dissipative operator $\K$, via a square root operation~\cite{Reina2015a}.
We remark that in this general setting, $\rho$ is a state variable, not necessarily a density. This shows the potential generality
of the approach, as~\eqref{eq:entropy_gradient_flow} is the dissipative (non-conservative) part of the GENERIC setting (General
Equations for Non-Equilibrium Reversible-Irreversible Coupling)~\cite{Ottinger2005a}.

The feasibility study presented here opens the door to many future investigations. In particular, the method could be extended to
more complex situations, including, multi-component systems or other transport phenomena. Furthermore, the sequential version,
sketched in Subsection~\ref{sec:Altern-meth-sequ}, is potentially promising for experimental data and deserves further
analysis. Establishing a rigorous theory is likely to be a demanding task, as for other methods for the determination of transport
coefficients. Indeed, in general key assumptions are known rigorously only in few cases. For example, mean square displacement
methods rely on the observation that a tagged particle behaves under diffusive rescaling like a Brownian motion with diffusion
matrix $\sigma$. While this proves very successful in a wide variety of cases, rigorous proofs are known only either in some
equilibrium situations or for the zero-range process out of equilibrium~\cite{Jara2013a}, one of test cases studied with the
method proposed here.

To conclude, we note that the method can provide insight in the continuum behaviour of particle systems whose
coarse-grained description is currently unknown, as shown in the following example.

\subsection{Outlook: Kawasaki dynamics}
\label{sec:Outl-Kawasaki-dynam}

\emph{Kawasaki dynamics} is a stochastic process used to model the evolution of surfaces, where the total mass of the substance
surrounded by the surface is conserved (see, for example,~\cite[Chapter~18]{Bovier2015a}). One version of this process can be
described as follows.  We assume the surface can be described as a graph, such that the height at lattice position $X$ is
$\eta(X)$. The following dynamics is a modification of the nearest neighbour Ising model with Kawasaki dynamics, which, unlike the
original model, preserves the graph property under evolution. Several particles can occupy a site and jump to a neighbouring
position with the rate based on the overall ``energy'' $H$ of the system, given by
\begin{equation}
  \label{eq:H_Kawa}
  H(\eta)=\frac{1}{2} \sum_{X} \sum_{\abs{X-\tilde X} = 1}
  (\eta(\tilde X )-\eta(X))^{2}.
\end{equation}
The rate for a particle at site $X$ to jump to $\tilde X$, thus changing the state from $\eta$ to $\eta^{X, \tilde X}$, is
\begin{equation*}
  g_{X\rightarrow \tilde X}(\eta)
  :=\begin{cases}
  \frac{e^{H(\eta^{X, \tilde X})}}{e^{H(\eta^{X, \tilde X })}+e^{H(\eta)}} & \text{if }\eta(X)>0 \text{ and } \abs{\tilde X -X} = 1 \\
    0 & \text{otherwise}
  \end{cases}.
\end{equation*}
Note that although $H$ depends on the whole state, the jump rate at a site only depends on the occupation numbers at the two sites
involved in the jump, and their neighbours. To our knowledge, neither the entropy $\S$ nor the hydrodynamic limit have been
rigorously established for this version of Kawasaki dynamics (for a related, simpler, model, the hydrodynamic limit can be
established~\cite{Nishikawa2017a}). The application of our method to this process is thus speculative. Yet, the governing metric
is expected to be an $H^{-1}$ metric~\cite{Grunewald2009a,Nishikawa2017a}.

\begin{figure}
  \centerline{\input{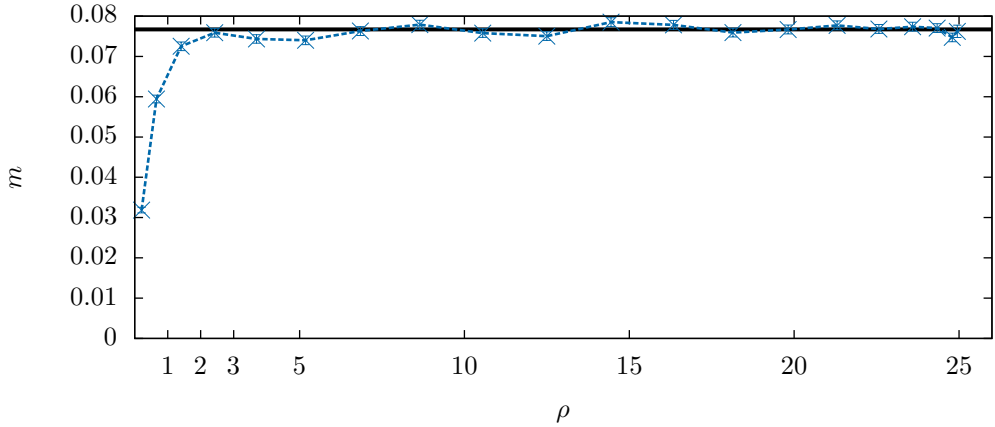}}
  \caption{Results for the Kawasaki type dynamics. Depicted is only the mobility $\m$, as the diffusivity $D$ would require the
    knowledge of the entropy of this process. The black, horizontal line at about $0.077$ is the mean of the data for large
    $\rho$.  The measurement is based on $a_1=40$, and an initial profile
    $\eta(t_{\ini}L^2,X) = 12.5\cdot\left(1-\cos\left(2\pi \frac{X}{\L}\right)\right)$ is chosen, to better capture the
    non-constant behaviour close to $\rho=0$. The other parameters are the default ones described in
    Subsection~\ref{sec:Defa-choice-parame}.}
  \label{fig:Kaw_m}
\end{figure}

While in principle the entropy $\S$ and the mobility $\m$ could depend on $\rho$ and $\nabla \rho$, a dependency of $m$ on $\rho$
is not expected. Such a conclusion is drawn from the fact that the Hamiltonian~\eqref{eq:H_Kawa} is invariant under additions to
$\eta$, and $\rho$ is the expectation of $\eta$.  We show in Figure~\ref{fig:Kaw_m} supporting evidence for this independence,
thus suggesting that the macroscopic evolution is governed by a norm independent of $\rho$, such as the $H^{-1}$ norm. We point
out that while the independence on $\rho$ is easily observed for high densities, the computational method presented here fails for
low densities in this test case. This is not surprising, as one would expect the constant profile (in $\rho$) to appear as a
result of cancellations, which might not be captured correctly in the numerical computations for small values of $\rho$. This
error is somewhat artificial, as only large densities are a good representation of the graph of the surface whose evolution
Kawasaki dynamics aims to study. 

\appendix

\section{Entropy gradient flow and the Wasserstein metric}
\label{sec:Entr-grad-flow}

The linear diffusion equation can be written as
\begin{equation}
  \label{eq:diff-linear}
  \partial_t \rho = \Delta \rho
  = -\div( \rho \nabla \DS(\rho)) =: \K(\rho) \DS (\rho) , 
\end{equation}
where $\S(\rho) = -\int \rho \log(\rho) \df x$ is the \emph{Boltzmann entropy} in dimensionless units; $\K$ is the operator
$\K(\rho) \xi = -\div(\rho \nabla \xi)$; and $\DS$ is the variational derivative, $\DS = -\log(\rho) - 1$. While the previous
identity is trivial to verify, the meaning of the term on the right-hand side goes much deeper. Namely, it can be shown that
$\K(\rho) \DS (\rho)$ is the steepest ascent of the entropy in a geometry associated to $\K$, the so-called \emph{Wasserstein
  geometry}~\cite{Jordan1998a}. We sketch some core results forming the background of this article, in particular the gradient
ascent in the Wasserstein geometry, for~\eqref{eq:diff-linear} on $\R^d$. It can be seen that natural setting
for~\eqref{eq:diff-linear} is $\rho \in P_2(\R^d)$, the space of probability measures with finite second moments. This space can
be equipped with the so-called \emph{Wasserstein metric}; a result by Benamou and Brenier~\cite{Benamou2000a} characterises this
metric in variational form,
\begin{equation}
  \label{eq:BB}
  d(\rho_0,\rho_1)^2 = \inf \int_0^1 \int_{\R^d} \rho(x,t) \abs{v(x,t)}^2 \df x \df t,  
\end{equation}
with (pathwise) minimisation over densities $\rho$ and velocities $v$ satisfying
\begin{equation}
  \label{eq:BB-cont}
  \partial_t \rho + \div (\rho v) = 0 . 
\end{equation}
Note that~\eqref{eq:BB} is a problem of optimal transport. Indeed, one can visualise~\eqref{eq:BB} as the cost of moving mass from
$\rho_0$ to $\rho_1$. The continuity equation ensures that mass is conserved along the transport; the distance is the optimal one,
i.e., the one that minimises the cost functional on the right-hand side of~\eqref{eq:BB}. Benamou and Brenier also show that the
velocity field is in fact a gradient, $v = \nabla \Psi$. Then, the norm~\eqref{eq:BB} gives formally rise to an inner product,
\begin{equation}
  \label{eq:wass-ip}
  (s_1, s_2)_{\K^{-1}} := \int \rho \nabla \Psi_1 \nabla \Psi_2 \df x , 
\end{equation}
where $s_j = -\div(\rho \nabla \Psi_j)$ for $j = 1, 2$. 

We are now in a position to see the Wasserstein gradient flow structure. Namely, a \emph{gradient flow} of a functional $\S$ in a
geometry given by an inner product is by definition an evolution of the kind that
\begin{equation}
  \label{eq:gf-general}
  (\partial_t \rho, s_2)_{\K^{-1}} =  \int \DS s_2 \df x
\end{equation}
for all suitable test functions $s_2$ with $s_2 = -\div (\rho \nabla \Psi_2)$. Here, this gives
 \begin{align*}
   \int_{\R^d} \rho \nabla \Psi_1 \nabla \Psi_2 \df x &  = - \int_{\R^d} (\log(\rho) + 1 )s_2 \df x  \\
   \intertext{with $\partial_t \rho = -\div(\rho \nabla \Psi_1)$ and $s_2 = -\div(\rho \nabla \Psi_2)$. An integration by parts gives}
   \int_{\R^d}  (\div)^{-1} (\partial_t \rho ) \nabla \Psi_2 \df x      
                                   & = \int_{\R^d} \nabla \rho \nabla \Psi_2 \df x , \\
   \intertext{from which we obtain by integrating by parts once more}
   \int_{\R^d}  \partial_t \rho \Psi_2 \df x  
                                  & = \int_{\R^d} \Delta \rho \Psi_2 \df x. 
\end{align*}
This is the weak form of the diffusion equation~\eqref{eq:diff-linear}.

For the zero range process discussed in Subsection~\ref{sec:Zero-range-process}, one can analogously define a gradient flow
structure. We refer the reader to~\cite{Dirr2016a}.

\vskip6pt

\paragraph{Acknowledgements} All authors thank the Leverhulme Trust for its support via grant RPG-2013-261. JZ gratefully
acknowledges funding by a Royal Society Wolfson Research Merit Award.  ND gratefully acknowledges funding by the EPSRC through
project EP/M028607/1. PE thanks Cardiff University through the International Collaboration Seedcorn Fund to visit CR. CR further
thanks the University Research Foundation at Penn. We thank Rob L.~Jack and Tim Rogers for helpful discussions and the reviewers
for their comments.




\def\cprime{$'$} \def\cprime{$'$} \def\cprime{$'$}
  \def\polhk#1{\setbox0=\hbox{#1}{\ooalign{\hidewidth
  \lower1.5ex\hbox{`}\hidewidth\crcr\unhbox0}}} \def\cprime{$'$}
  \def\cprime{$'$}

%
%
%
%
%
%
%
%
%

\begin{thebibliography}{99}

\bibitem{Frenkel2002a}
Frenkel D, Smit B. 2002 {\em Understanding Molecular Simulation} vol.~1{\em
  Computational Science Series}.
Orlando, FL, USA: Academic Press, Inc. second edition.

\bibitem{Keffer2005a}
Keffer DJ, Gao CY, Edwards BJ. 2005  On the Relationship between {F}ickian
  Diffusivities at the Continuum and Molecular Levels. {\em J. Phys. Chem. B}
  \textbf{109}, 5279--5288.
PMID: 16863195.

\bibitem{Allen1989a}
Allen MP, Tildesley DJ. 1989 {\em Computer Simulation of Liquids}.
New York, NY, USA: Clarendon Press.

\bibitem{Tuckerman2010a}
Tuckerman ME. 2010 {\em Statistical mechanics: theory and molecular
  simulation}.
Oxford Graduate Texts. Oxford University Press, Oxford.

\bibitem{Evans2007a}
Evans DJ, Morriss GP. 2007 {\em Statistical Mechanics of Nonequilibrium
  Liquids}.
ANU Press.

\bibitem{Jordan1998a}
Jordan R, Kinderlehrer D, Otto F. 1998  The variational formulation of the
  {F}okker-{P}lanck equation. {\em SIAM J. Math. Anal.} \textbf{29}, 1--17
  (electronic).

\bibitem{Reina2015a}
Reina C, Zimmer J. 2015  Entropy production and the geometry of dissipative
  evolution equations. {\em Phys. Rev. E} \textbf{92}, 052117, 7.

\bibitem{Ottinger2005a}
\"Ottinger HC. 2005 {\em Beyond equilibrium thermodynamics}.
Wiley Online Library.

\bibitem{Mielke2011b}
Mielke A. 2011  Formulation of thermoelastic dissipative material behavior
  using {GENERIC}. {\em Contin. Mech. Thermodyn.} \textbf{23}, 233--256.

\bibitem{Dean1996a}
Dean DS. 1996  Langevin equation for the density of a system of interacting
  {L}angevin processes. {\em Journal of Physics A: Mathematical and General}
  \textbf{29}, L613.

\bibitem{Eyink1996a}
Eyink GL, Lebowitz JL, Spohn H. 1996  Hydrodynamics and fluctuations outside of
  local equilibrium: driven diffusive systems. {\em J. Statist. Phys.}
  \textbf{83}, 385--472.

\bibitem{Jack2014a}
Jack RL, Zimmer J. 2014  Geometrical interpretation of fluctuating
  hydrodynamics in diffusive systems. {\em J. Phys. A} \textbf{47}, 485001, 17.

\bibitem{Hurtado2013a}
Hurtado PI, Lasanta A, Prados A. 2013  Typical and rare fluctuations in
  nonlinear driven diffusive systems with dissipation. {\em Phys. Rev. E}
  \textbf{88}, 022110.

\bibitem{Renesse2009a}
von Renesse MK, Sturm KT. 2009  Entropic measure and {W}asserstein diffusion.
  {\em Ann. Probab.} \textbf{37}, 1114--1191.

\bibitem{Kipnis1999a}
Kipnis C, Landim C. 1999 {\em Scaling limits of interacting particle systems}
  vol. 320{\em Grundlehren der Mathematischen Wissenschaften [Fundamental
  Principles of Mathematical Sciences]}.
Berlin: Springer-Verlag.

\bibitem{Ferrari1988a}
Ferrari PA, Presutti E, Vares ME. 1988  Non equilibrium fluctuations for a zero
  range process. {\em Ann. Inst. H. Poincar\'e Probab. Statist.} \textbf{24},
  237--268.

\bibitem{Landim2008a}
Landim C, Milan{\'e}s A, Olla S. 2008  Stationary and nonequilibrium
  fluctuations in boundary driven exclusion processes. {\em Markov Process.
  Related Fields} \textbf{14}, 165--184.

\bibitem{Barndorff-Nielsen2002a}
Barndorff-Nielsen OE, Shephard N. 2002  Econometric analysis of realized
  volatility and its use in estimating stochastic volatility models. {\em J. R.
  Stat. Soc. Ser. B Stat. Methodol.} \textbf{64}, 253--280.

\bibitem{Oksendal2003a}
{\O}ksendal B. 2003 {\em Stochastic differential equations}.
Universitext. Springer-Verlag, Berlin sixth edition.
An introduction with applications.

\bibitem{Guckenheimer1990a}
Guckenheimer J, Holmes P. 1990 {\em Nonlinear oscillations, dynamical systems,
  and bifurcations of vector fields} vol.~42{\em Applied Mathematical
  Sciences}.
Springer-Verlag, New York.
Revised and corrected reprint of the 1983 original.

\bibitem{Dixit2016a}
Dixit UJ. 2016 {\em Examples in parametric inference with {R}}.
Springer, Singapore.

\bibitem{Bortz1975a}
Bortz AB, Kalos MH, Lebowitz JL. 1975  A new algorithm for {M}onte {C}arlo
  simulation of {I}sing spin systems. {\em J. Computational Phys.} \textbf{17},
  10--18.

\bibitem{Grosskinsky2003a}
Grosskinsky S, Sch{\"u}tz GM, Spohn H. 2003  Condensation in the zero range
  process: stationary and dynamical properties. {\em J. Statist. Phys.}
  \textbf{113}, 389--410.

\bibitem{Abramowitz1964a}
Abramowitz M, Stegun IA. 1964 {\em Handbook of mathematical functions with
  formulas, graphs, and mathematical tables} vol.~55{\em National Bureau of
  Standards Applied Mathematics Series}.
For sale by the Superintendent of Documents, U.S. Government Printing Office,
  Washington, D.C.

\bibitem{Adams2013a}
Adams S, Dirr N, Peletier M, Zimmer J. 2013  Large deviations and gradient
  flows. {\em Philos. Trans. R. Soc. Lond. Ser. A Math. Phys. Eng. Sci.}
  \textbf{371}, 20120341, 17.

\bibitem{Arratia1983a}
Arratia R. 1983  The motion of a tagged particle in the simple symmetric
  exclusion system on {${\bf Z}$}. {\em Ann. Probab.} \textbf{11}, 362--373.

\bibitem{Eyink1990a}
Eyink GL. 1990  Dissipation and large thermodynamic fluctuations. {\em J.
  Statist. Phys.} \textbf{61}, 533--572.

\bibitem{Jara2013a}
Jara M, Landim C, Sethuraman S. 2013  Nonequilibrium fluctuations for a tagged
  particle in one-dimensional sublinear zero-range processes. {\em Ann. Inst.
  Henri Poincar\'e Probab. Stat.} \textbf{49}, 611--637.

\bibitem{Bovier2015a}
Bovier A, den Hollander F. 2015 {\em Metastability} vol. 351{\em Grundlehren
  der Mathematischen Wissenschaften [Fundamental Principles of Mathematical
  Sciences]}.
Springer, Cham.
A potential-theoretic approach.

\bibitem{Nishikawa2017a}
Nishikawa T. 2017  Hydrodynamic limit for the {G}inzburg-{L}andau
  {$\nabla\phi$} interface model with a conservation law and {D}irichlet
  boundary conditions. {\em Stochastic Process. Appl.} \textbf{127}, 228--272.

\bibitem{Grunewald2009a}
Grunewald N, Otto F, Villani C, Westdickenberg MG. 2009  A two-scale approach
  to logarithmic {S}obolev inequalities and the hydrodynamic limit. {\em Ann.
  Inst. Henri Poincar\'e Probab. Stat.} \textbf{45}, 302--351.

\bibitem{Benamou2000a}
Benamou JD, Brenier Y. 2000  A computational fluid mechanics solution to the
  {M}onge-{K}antorovich mass transfer problem. {\em Numer. Math.} \textbf{84},
  375--393.

\bibitem{Dirr2016a}
Dirr N, Stamatakis M, Zimmer J. 2016  Entropic and gradient flow formulations
  for nonlinear diffusion. {\em J. Math. Phys.} \textbf{57}, 081505, 13.

\end{thebibliography}

\end{document}